\documentclass[preprint]{aastex}
\pdfoutput=1
\newcommand{\co}{$^{13}$CO}
\newcommand{\kms}{km~s$^{-1}$}
\newcommand{\msun}{$M_{\sun}$}
\newcommand{\cc}{cm$^{-3}$}
\newcommand{\hii}{\ion{H}{2}}
\newcommand{\sst}{{\it Spitzer Space Telescope}}
\newcommand{\spit}{{\it Spitzer}}
\newcommand{\hso}{{\it Herschel Space Observatory}}
\newcommand{\her}{{\it Herschel}}
\newcommand{\sfr}{G38.9-0.4}
\newcommand{\err}{$\pm$}
\newcommand{\gcm}{g~cm$^{-2}$}
\newcommand{\mpc}{$M_{\sun}$~pc$^{-2}$}
\newcommand{\av}{$A_{\mathrm{V}}$}
\usepackage{subfigure}

\begin{document}
\title{THE INTERSTELLAR BUBBLES OF G38.9-0.4 AND THE IMPACT OF STELLAR FEEDBACK ON STAR FORMATION}

\author{Michael J. Alexander}
\email{malexan9@uwyo.edu}
\author{Henry A. Kobulnicky}
\email{chipk@uwyo.edu}
\affil{Department of Physics \&\ Astronomy, University of Wyoming,1000 E. University Avenue, Laramie, WY 82071, USA}
\author{Charles R. Kerton}
\email{kerton@iastate.edu}
\affil{Department of Physics \&\ Astronomy, Iowa State University, Ames, IA 50011, USA}
\and
\author{Kim Arvidsson}
\affil{Adler Planetarium, 1300 S Lake Shore Drive, Chicago, IL 60605, USA}
\email{karvidsson@adlerplanetarium.org}

\begin{abstract} 

We present a study of the star formation (SF) region G38.9-0.4 using publicly available multiwavelength Galactic Plane surveys from ground- and space-based observatories. This region is composed of four bright mid-IR bubbles and numerous infrared dark clouds. Two bubbles, N~74 and N~75, each host a star cluster anchored by a single O9.5V star. We identified 162 young stellar objects (YSOs) and classify 54 as stage I, 7 stage II, 6  stage III, and 32 ambiguous. We do not detect the classical signposts of triggered SF, i.e., star-forming pillars or YSOs embedded within bubble rims. We conclude that feedback-triggered SF has not occurred in G38.9-0.4. The YSOs are preferentially coincident with infrared dark clouds. This leads to a strong correlation between areal YSO mass surface density and gas mass surface density with a power law slope near 1.3, which closely matches the Schmidt-Kennicutt Law. The correlation is similar inside and outside the bubbles and may mean that the SF efficiency is neither enhanced nor supressed in regions potentially influenced by stellar feedback. This suggests that gas density, regardless of how it is collected, is a more important driver of SF than stellar feedback. Larger studies should be able to quantify the fraction of all SF that is feedback-triggered by determining the fraction SF, feedback-compressed gas surrounding \ion{H}{2} regions relative to that already present in molecular clouds. 

\end{abstract}


\section{Introduction}
\citet{el77} proposed a mechanism where massive ($>$8 \msun) OB stars could trigger a new wave of star formation (SF) that would not have occurred otherwise. This `collect-and-collapse' (CC) model showed that the expansion of an \hii\ region into a uniform interstellar medium (ISM) can sweep up shells dense enough to collapse and form stars. A second proposed triggering mechanism is through radiation-driven implosion (RDI; \citealt[][]{be89}). \citet{le94} demonstrated that ionizing ultraviolet (UV) radiation impinging upon a cloud can compress clumps of ISM to induce collapse. In their simulation, UV photons were directed at a preexisting, gravitationally stable core ($n = 1550$ \cc) embedded in neutral ISM ($n = 10$ \cc). The resulting gas formed pillar-like structures with gravitationally unstable, SF cores at the tips, similar to what is observed in many star forming regions (SFRs) \citep{re04,in07,bi10}. Hydrodynamic simulations by \citet{gr10} investigated the effect of radiation on molecular clouds with turbulent gas. Their ``radiative round-up" (RR) model produced star forming pillars from the turbulent gas with distinct differences in line-of-sight gas profiles compared to RDI. 

The observational signatures of triggered SF should differ depending on the dominant mode, but in general, evidence for triggered SF will show patterns in the distribution of molecular gas, the distribution of YSOs, and the ages of YSOs relative to the \hii\ regions. CC is a relatively slow process that takes on the order of a few Myr, which creates {\it massive} clumps that can collapse to form massive stars or star clusters \citep{de05}. The expansion of the \hii\ region triggers SF, which implies that YSOs should be found within, or adjacent to, the clumps that form on the periphery of the bubbles \citep{de08,wa08}. The CC model also predicts that (in a uniform medium) the collected shell will fragment at regular intervals and lead to uniformly distributed SF clumps \citep{de05}. As the ionization front pushes outward, new stars should continue to form (provided there is enough molecular material) and leave behind older, more evolved YSOs. Observationally, this predicts a YSO age gradient that can be used to identify cases of triggered SF. On the other hand, RDI can form young stellar objects (YSOs) in just a few tenths of a Myr, and the stars are intermediate mass, 1--7 \msun\ \citep{le94}. RR differs from typical CC models which assume a static, uniform medium, and unlike RDI models, does not require pre-existing cores of molecular material. The RR model forms sub-solar mass protostars on timescales similar to RDI ($\approx$few $10^{5}$ yr). The stars formed in both CC and RR should have near-zero velocities relative to the expanding \hii\ shell because they formed out of ISM already in motion. In RDI, the cores are formed before the expansion of the \hii\ region, and therefore the resultant YSOs may have non-zero  velocities with respect to the \hii\ shell. The uniqueness of individual SFRs make it difficult to attribute specific features to a given mode of triggered SF (CC, RDI, or RR), however, the observational signatures mentioned above can be used to identify potential sites of triggered SF.

This paper is the first in a series that will investigate triggered SF over a variety of Galactic environments. Our goal is to examine, in detail, the properties of the SFRs themselves to determine the input energy required for triggered SF to be an effective and efficient mechanism. This requires knowledge of the triggering and triggered populations, as well as the molecular cloud out of which they formed. To answer these questions we have gathered survey data from numerous sources, from the near-IR to the radio, to investigate the Galactic  SFR \sfr. This region was chosen because it is relatively nearby (2.7 kpc) and lies along an unconfused line of sight. It also appears to host bubbles at different evolutionary stages and may provide insight into the growth of \hii\ regions and the potential for triggered SF.

\begin{figure}
\includegraphics[width=18cm]{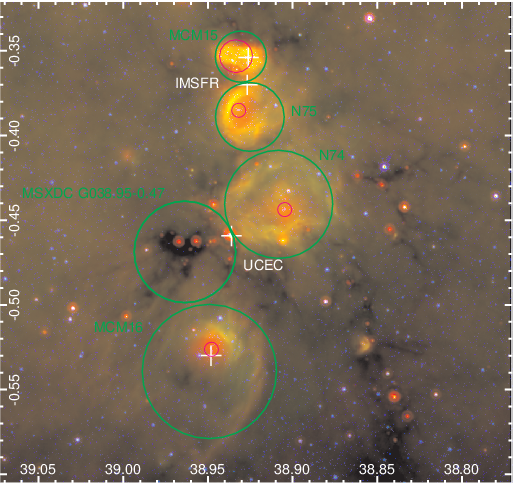}
\caption{Three-color image of the \sfr\ region. Blue is UKIDSS K, green is IRAC 8 \micron, and red is MIPS 24 \micron. The specific regions discussed in the paper are labeled. Green circles indicate the areas used for gas mass and radio continuum calculations, while magenta circles indicate the apertures used for cluster identification. The upper-most and lower-most white crosses mark the center location of the candidate cluster in \citep{me05}, while the inner crosses mark the location of the IMSFR IRAS source and UCEC. \label{rgb}}
\end{figure}

Figure~\ref{rgb} shows a three-color image of \sfr\ with UKIDSS K-band in blue, {\it Spitzer} IRAC 8 \micron\ in green, and {\it Spitzer} MIPS 24 \micron\ in red. The green circles and text mark the main features of the region and indicate the apertures used for estimating the total gas mass and radio continuum flux associated with each feature in subsequent sections. This region contains two prominent IR bubbles (N~74 and N~75; \citealt[][]{ch06}), two cataloged star clusters (\#15 and \#16, hereafter MCM~15 and MCM~16; \citealt[][]{me05}), an intermediate-mass SFR (IMSFR, IRAS 19012+0505; \citealt[][]{ar10}), an infrared dark cloud (IRDC) MSXDC G038.95-0.47 (\citealt[][]{ra06}; hereafter just IRDC), and an ``ultracompact embedded cluster" (UCEC \#8; \citealt[][]{al12}). Magenta circles outline the areas used to find possible stellar clusters contained within the bubbles. The white crosses indicate the center positions of the features mentioned above as they are given in the cited references. Due to the IRAS beam size, the IMSFR is a blend between the mid-IR emission from MCM~15 and the N~75 bubble. The UCEC is the bright point-like feature located between the IRDC and N~74.

In Section~\ref{ismsec}, we characterize the nature of the ISM and present evidence for the presence massive stars. Section~\ref{ysosec} describes the method for finding YSOs. Section~\ref{fieldsec} assesses the main-sequence (MS) stellar content of the region. Section~\ref{regionsec} is a breakdown of individual subregions. Section~\ref{discsec} is the discussion, including possible relationships between individual YSO masses and gas surface density, and between YSO mass surface density and gas mass surface density. We also discuss the YSO mass function in relation to other studies and the possibility of triggered SF. A summary of conclusions is presented in Section~\ref{concsec}.

\section{Interstellar Environment \label{ismsec}}
\subsection{Molecular Gas in \co}
\begin{figure}
\includegraphics[width=18cm]{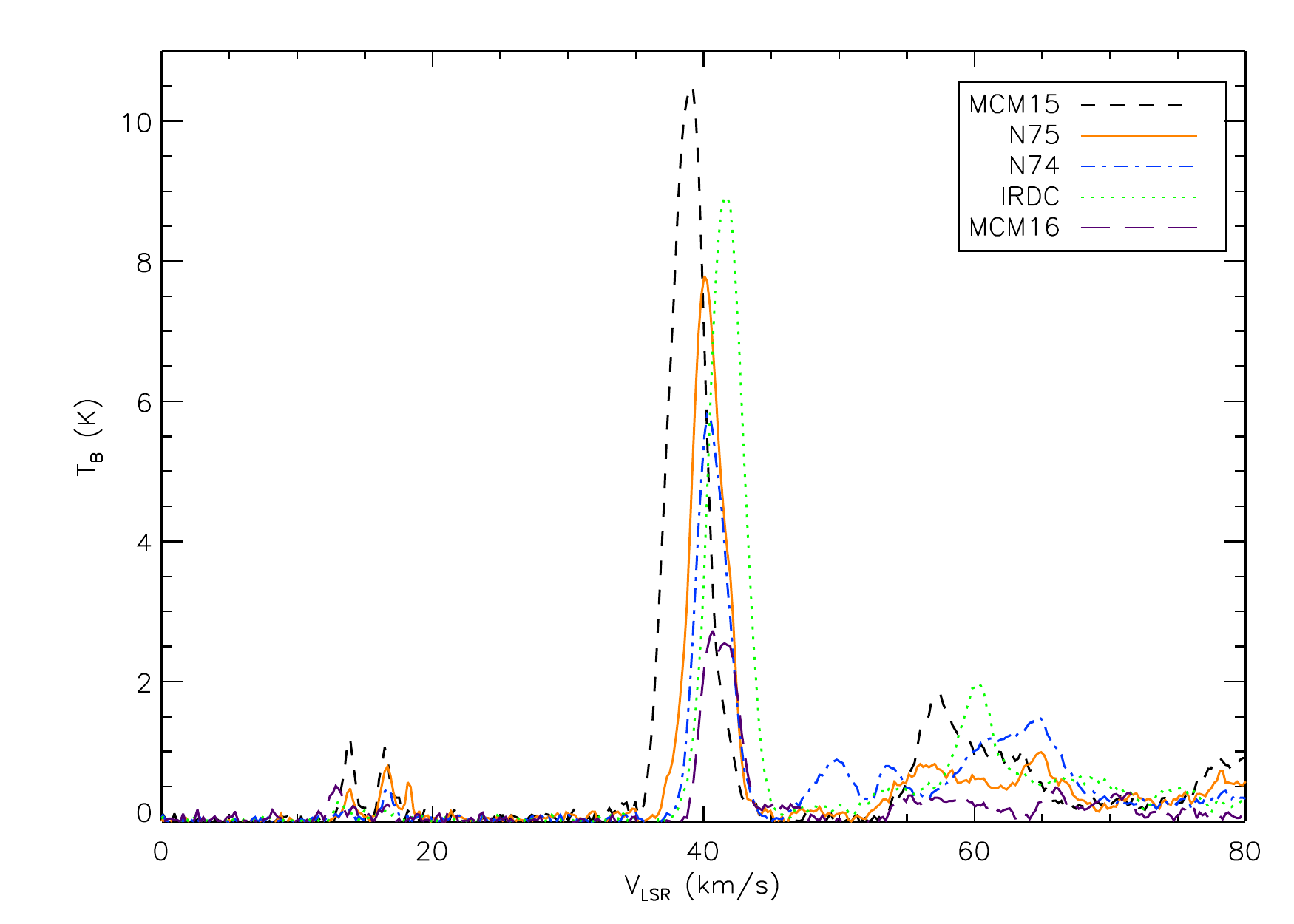}
\caption{T$_{B}$ versus $V_{\mathrm{LSR}}$ plot of \co\ for the main regions of \sfr. Data for MCM~15 is shown as the black short-dashed line, N~75 the orange solid line, N~74 the blue dot-dashed line, the IRDC the green dotted line, and MCM~16 the magenta long-dashed line. Associated molecular gas lies in the range of 35--45 \kms, while peaks at higher velocities are foreground/background clouds with no apparent spatial correlation. The individual peaks show a slight shift towards higher velocities from North (MCM~15) to South (IRDC). The tangent velocity for this longitude is 79 \kms. \label{velspec}}
\end{figure}

One indicator of active and future SF is the presence of molecular gas. To characterize this component we use data from the Boston University Five College Radio Astronomy Observatory Galactic Ring Survey (BU-FCRAO GRS; \citealt{ja06}), which targeted the \co\ $J=1\rightarrow 0$ transition at 110.2 GHz with a beam FWHM of 46\arcsec and velocity resolution of 0.21 \kms. \co\ is much less abundant than H$_{2}$ or $^{12}$CO and thus remains optically thin at higher densities. We extracted spatially averaged velocity spectra from the GRS data cubes within N~74, N~75, cluster MCM~15, and the known IRDC. Figure~\ref{velspec} shows the 1D velocity spectrum for each region with MCM~15 as the black short-dashed line, N~75 as the orange solid line, N~74 as the blue dot-dashed line, the IRDC as the green dotted line, and MCM~16 as the magenta, long-dashed line.  The \co\ brightness temperature peaks between 5.6~K and 10~K over the velocity range $V_{\mathrm{LSR}}$ = 39--42 \kms\ for all four regions, indicating abundant, probably related, molecular material over a small velocity interval. The peak  velocities for each region increase slightly from north to south with a spread of about 3 \kms\ between the peaks. There is also low-level emission detected from 50~--~65 \kms\ for each region. There is no morphological similarity between this emission and the IR features so it is likely to be unaffiliated diffuse ISM.

\begin{figure}
\includegraphics[width=18cm]{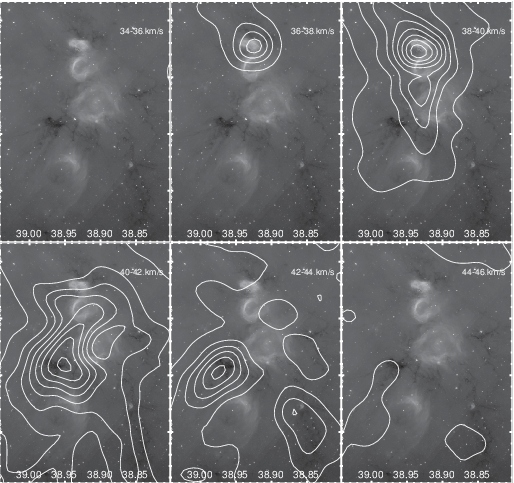}
\caption{Channel plot showing \co\ contours overlaid on an IRAC 8 \micron\ image. The contours run linearly from 0.5 to 10.5 K~\kms\ in steps of 1.25 K~\kms, and are averaged every 2 \kms\ over the range 34--46 \kms. The velocity transition is seen across the different regions. The maximum value in each panel is (from left-to-right, top-to-bottom) 0.4 K~\kms, 5.6 K~\kms, 12.0 K~\kms, 10.7 K~\kms, 8.0 K~\kms, and 1.8 K~\kms. \label{channel}}
\end{figure}

Figure~\ref{channel} is a six-panel \co\ channel map for the SFR from 35 to 45 \kms, averaged every 2 \kms, overlaid on a GLIMPSE 8 \micron\ image. There is obvious spatial agreement between the molecular gas and IR emission. There is a strong peak coincident with MCM~15 at 36 -- 38 \kms\ and a smooth transition across the image until another strong peak appears over the IRDC at 43 \kms. The 43 \kms\ panel also seems to show a \co\ hole over the center of N~74. This feature may be evidence for a massive star sweeping out molecular material. The smooth variation in \co\ velocity across this region suggests that all of these features are affiliated with the same molecular complex at a common distance. We adopt an average velocity of 41 \kms\ for the molecular material and obtain near and far kinematic distances of 2.7 and 10.5 kpc, using the Galactic rotation curve of \citet{cl85}. \citet{ra06} chose the near distance for the IRDC based on the assumption that it must be closer than the bright IR emission in order to appear in silouhette. Also, the UKIDSS images exhibit a noticeable decline in source counts coincident with IRDCs relative to the field at large. Near-IR source catalogs from the UKIRT Infrared Deep Sky Survey (UKIDSS; \citealt[][]{la07}) show 80 -- 160 sources per square arcminute coincident with the IRDCs, while the field away from the molecular shows typically shows over 400 sources per square arcminute. This indicates that a large fraction, up to 75\% of sources are being obscured at the location of the molecular cloud. If the molecular cloud and associated IR features were at the far distance of 10.5 kpc, then the cloud would be behind a majority of the field stars and any decline in source counts would be much less noticeable. Therefore, we adopt the near distance of 2.7 kpc for the entire region. The difference between the lowest and highest peak emission velocities for the individual regions in Figure~\ref{velspec} amounts less than 0.2 kpc in distance, which is less than the assumed error in kinematic distances which we take to be 15\%.  The observed velocity spread is consistent with observed internal velocity dispersions of molecular clouds of several \kms\ \citep{so87, ba11}. We estimate the column density of the \sfr\ molecular cloud from the GRS \co\ cube using the formula from \citet{si01},
\begin{equation}
N(^{13}CO) = 8.75\times10^{14}~T_{mb}\Delta v ~~\mathrm{cm^{-2} , }
\end{equation}
where $T_{mb}$ is the main beam brightness temperature in K and $\Delta v$ is the velocity range in \kms, for which we used 35--45 \kms. This equation assumes a \co\ excitation temperature of 10~K, which is consistent with the estimated average value for the cloud of 8.3 K \citep{ro10}. We then compute the total gas mass surface density, $\Sigma_{\mathrm{CO}}$, in \gcm\ from \co\ using the conversion factors R($^{12}$CO/\co) = 45 and X($^{12}$CO/H$_{2}$) = $8\times10^{-5}$ \citep{si01} and multiplying by $\mu m_{H}$, the mean molecular weight (2.72; \citealt[][]{ro10}) times the mass of neutral hydrogen. The total mass is the mass surface density summed over the physical size of an individual region, which was estimated using a distance of 2.7 kpc. The mass errors are from the flux estimation, the assumed distance, and the adopted temperature. Table~\ref{params} shows the results for the five main subregions and for the entire cloud. Column 1 gives the region name, columns 2 and 3 list the center coordinates, and column 4 is the nominal diameter of each subregion. The specified areas are used to calculate the total gas mass for each subregion. Column 5 gives the peak velocity for each subregion and column 7 gives the \co\ derived gas mass.

\subsection{Submillimeter Dust Continuum}
One of the key projects undertaken with the \hso\ was the \her\ Infrared Galactic Plane Survey, or Hi-GAL \citep{mo10a}. Hi-GAL covers the Galactic Plane at 70, 160, 250, 350, and 500 \micron\ from $\mid l\mid<60\degr$ and $\mid b\mid<1\degr$. The advantage of using Hi-GAL millimeter wavelength dust continuum over \co\ line emission to trace molecular gas is that carbon-based molecules deplete onto dust grains when densities exceed 10$^{4}$ \cc, while mm-continuum remains optically thin at the higher densities expected in IRDCs and SF cores \citep{ta04}. While the \co\ observations have the advantage of separating foreground and background emission to achieve a better estimate of total cloud mass, the sub-millimeter continuum should provide more accurate measurements in dense regions, such as IRDCs and SF clumps.

The Hi-GAL wavelength coverage spans the peak of thermal dust emission (15 K at $\sim$100 \micron), which allows for temperature fitting and more accurate surface density estimates. We made a dust temperature map by first aligning and resampling the \her\ maps to the 3\farcs2 pixel scale of the 70 $\mu$m maps using bi-linear interpolation. This approach has the advantage of perserving angular resolution and information in the short-wavelength maps but assumes that the emission at longer wavelengths varies smoothly over sub-pixel scales (14\arcsec\ pixel$^{-1}$ at 500 \micron). We then made integrated velocity maps of associated \co\ emission (35--45 \kms) and the total line-of-sight emission up to the tangent point velocity (0--79 \kms), and used the ratio of the two to determine the fraction of sub-mm emission associated with the SFR. This method makes the assumption that sub-millimeter emission scales linearly with \co\ emission, which may not be the case over the wide density range observed. The density where \co\ becomes optically thick and where it depletes onto dust grains will alter the ratio. As a result of these two effects, we may be underestimating the associated sub-mm flux, leading to lower masses and surface densities. This is only important in higher density regions, such as IRDCs, and should not affect typical diffuse ISM.

After subtracting possible contaminating foreground/background emission from the total in each band map, we used the IDL fitting routine MPFIT \citep{mar09} to fit the \her\ 5-band data, pixel-by-pixel, to a modified blackbody of the form:
\begin{equation}
I_{\nu} = B_{\nu}(T)~(1 - e^{\tau_{\nu}})~~~\mathrm{MJy~sr^{-1}~~,}
\end{equation}
\begin{equation}
\tau_{\nu} = \tau_{o}\left(\frac{\nu}{\nu_{o}}\right)^{\beta}~~\mathrm{,}
\end{equation}
where $B_{\nu}(T)$ is the Planck function, and $\tau_{o}$ is the optical depth at frequency $\nu_{o}$. For the fitting we chose $\nu_{o}$ = ($c$/250~\micron), $\beta$ = 1.75, and left $\tau_{o}$ and $T$ as free parameters. We allow the temperature, $T$, to range from 1 to 100 K and $\tau_{250}$ from 0 to 1. We also required data in at least three bands at any given pixel for be fit. Values for $\nu_{o}$ and $\beta$ were adopted from \citet{ra10} and \citet{bat11}.

\begin{figure}
\includegraphics[width=18cm]{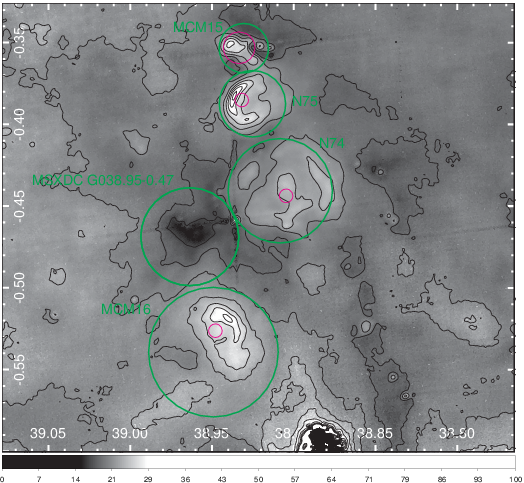}
\caption{Temperature map of the SFR. Green and magenta circles show the location of important features as in Figure~\ref{rgb}. Black contours indicate temperatures from 15 to 45 K, in steps of 3 K. The highest temperatures are associated with the bubble rims, while the coolest temperatures fall over IRDCs. The black region near $l = 38\fdg88$ at the bottom of the image is due to lack of data necessary to complete the temperature fitting. \label{temp}}
\end{figure}

Figure~\ref{temp} shows a grayscale image of the derived temperature map. Black contours outline 3 K steps from 15--45 K, while green and magenta circles outline the different sub-regions as before. We derive average temperatures for off-cloud regions of 19--22 K, which agrees well with typical Galactic cirrus \citep{bat11}. Fitted IRDC temperatures generally range from 14--22 K, and the bright rims of the bubbles have temperatures of 25--45 K. Computed average values for the subregions are listed in column 6 of Table~\ref{params}, these temperatures fit with the general picture that IRDCs contain cold SF cores, while the bright-rimmed clouds are \hii\ region shells being heated by the winds and UV photons from massive stars.

Our procedures and results are similar to those from \citet{ra10}, except they use a variable $\beta$. We performed a trial fit with $\beta$ variable from 0.5--2.5, and the final temperatures were generally within 20\%. The largest changes were in the IRDCs where $\beta$ dropped to 0.6--1.0 and temperature increased from 15 K to 20 K. The bright rims were also fit to lower $\beta$ in the range of 1.5, which increased the temperatures by 3--5 K. However, with at most 5 data photometric points, we deemed it necessary to reduce the number of free parameters and adopt a nominal fixed value for $\beta$.

From the temperature map, we calculated the pixel-by-pixel total gas mass surface density with
\begin{equation}
\Sigma_{\mathrm{500}} = \frac{I_{\nu}}{B_{\nu}(T)\kappa_{\nu}x_{dust}}~~~\mathrm{g~cm^{-2},}
\end{equation}
where I$_{\nu}$ is the observed brightness at 500 $\mu$m in MJy~sr$^{-1}$, $\kappa_{\nu}$ is the frequency-dependent opacity of 5.04 cm$^{2}$~g$^{-1}$, $x_{dust}$ is the dust-to-gas ratio of 0.01, and B$_{\nu}$(T) is the Planck function in MJy~sr$^{-1}$. The opacity was adopted from the tables of \citet{os94} assuming thin ice mantles at a density of $10^{6}$~cm$^{-2}$ after $10^{5}$ yr. The surface densities are then multiplied by the area to arrive at a total gas mass, which for the IRDC is 1386$\pm$418 \msun. \citet{ra06}, using 1.2 mm continuum data from the IRAM 30 meter and a dust temperature of 15 K, derived a mass of 1792 \msun, in rough agreement with our estimations from both \co\ and 500 \micron\ measurements. Full results are listed in column 8 of Table~\ref{params}.  We calculated the surface density for each of the five bands and found similar results for 170, 250, 350, and 500 \micron, but 70 \micron\ varied significantly. We adopted the 500 \micron\ values as this wavelength has the lowest optical depth and is likely to be the most reliable tracer of dense gas.

\subsection{Ionized gas as seen by radio continuum}
Figure~\ref{radio} is a three-color image with 500 \micron\ in blue, 8 \micron\ in green, and 24 \micron\ in red. White contours outline the 1.4 GHz radio continuum from VLA Galactic Plane Survey (VGPS; \citealt[][]{st06}). The minimum level is the mean background temperature plus two times the standard deviation, which is 13.15 K + 2(0.43) K = 14.01 K. Subsequent contours are 4, 6, 8, 10 and 12$\sigma$ above the mean background temperature. The continuum images from the VGPS have 1\arcmin\ resolution. The figure shows strong sub-mm emission from cold dust coincident with the IRDCs and on the peripheries of the bubbles. The radio emission peaks inside the two bubbles N~74 and N~75 and exhibits a lesser peak near MCM~16. Fluxes were extracted for N~74, N~75, and MCM~16 using IMVIEW, but there was no detection for MCM~15. Considering, the weak emission from MCM~16, it may be appropriate to treat the flux as an upper limit. The radio emission within the bubbles is most likely free-free emission from \hii, so we estimate the number of ionizing photons necessary to produce the 1.4 GHz flux by:
\begin{equation}
N_{\mathrm{Lyc}} = 7.5\times 10^{46}~F_{\nu}~\nu^{0.1}~d^{2}~T_{e}^{-0.45}~~~\mathrm{ph~s^{-1},}
\end{equation}
from \citet{ma76}, where F$_{\nu}$ is the flux density in Jy, $\nu$ is the frequency in GHz, d is the distance in kpc, and T$_{e}$ is the electron temperature in 10$^{4}$ K. We adopt 2.7 kpc for the distance, 8800 K for the electron temperature \citep{balser2011}, and the frequency is of the VGPS is 1.42 GHz. The results are summarized in Table~\ref{params}, where column 9 is the measured flux, column 10 is the computed log$_{10}$(N$_{\mathrm{Lyc}}$) ph~s$^{-1}$ and column 11 is the O9.5V equivalent (N$_{\mathrm{Lyc}}$(O9.5V) = 10$^{47.88}$ ph~s$^{-1}$; \citealt[][]{ma05}). We find a UV flux of N$_{Lyc}$ = 10$^{46.06}$ ph~s$^{-1}$ for N~74, in agreement with the estimate of N$_{Lyc}$ = 10$^{46.03}$ ph~s$^{-1}$ from \citet{be10}. Despite these values for the Lyman continuum flux, it is still likely that the bubbles N~74 and N~75 each contain a late-O star (see Section~\ref{regionsec}). Indeed, \citet{wa08} investigated the O star content of three prominent bubbles (N10, N12, and N49 from \citet{ch06}) and found that the radio flux underestimated the ionizing flux by a factor of two when compared with estimates based on the known spectral type through optical spectroscopy. This is likely because a significant fraction of UV photons are being absorbed by dust and heating the grains, as evidenced by the bright 24 \micron\ emission within IR bright bubbles. Additional losses may be incurred if UV photons escape the bubbles without being absorbed; therefore radio estimations of Lyman continuum emission should be treated as lower limits. Using a single O9.5V star as the estimated O star content of N~74 and N~75, we can calculate the mechanical wind luminosity, $L_{w}$,
\begin{equation}
L_{w} = 0.5~\dot{m_{w}}~v_{w}^{2}~~\mathrm{erg~s}^{-1}\mathrm{,} \label{lw}
\end{equation}
where $\dot{m}_{w}$ is the mass-loss rate and $v_{w}$ is the wind velocity. This yields $10^{32.34}$ ergs~s$^{-1}$ for nominal values of $10^{-9.5}~$M$_{\sun}~$yr$^{-1}$ and 1500 \kms\ for the mass loss rate and wind velocity of an O9.5V star \citep{marco09}. We then calculate the \hii\ region expansion radius as a function of time from \citet{we77} as,
\begin{equation}
R = 27~L_{w}^{0.2}~n_{o}^{-0.2}~t_{6}^{0.6}~~\mathrm{pc,}
\end{equation}
where $L_{w}$ is the wind luminosity in units of $10^{36}$ erg~s$^{-1}$, $n_{o}$ is the ambient ISM number density in 10$^{3}$ \cc, and t$_{6}$ is the expansion time in Myr. Figure~\ref{age} plots the radius versus age of expanding \hii\ regions for three different initial average density values (10$^{3}$, 10$^{4}$, and 10$^{5}$ \cc). The horizontal lines indicate the nominal radii of 0.65 pc and 1.1 pc for the N~75 and N~74 bubbles, respectively. The lower density (10$^{3}$ \cc) is typical of molecular clouds and higher values are approaching the density of IRDCs and cold molecular cores. For initial densities between 10$^{3}$ -- 10$^{3}$ \cc, we find a range of 0.4 -- 0.8 Myr for the expansion age of N~75 and 1.0 -- 2.2 Myr for N~74. RDI forms stars in a few $10^{5}$ yr after the ionization front has reached the existing clumps and CC may take a few Myr of \hii\ region expansion for SF to begin. The apparent youth of N~75 may rule out SF via CC or RDI, while N~74 may be old enough for either process to have formed stars, depending upon the intial conditions

\section{YSO Identification and Properties \label{ysosec}}
We identified YSOs by fitting their observed near-IR (NIR) and mid-IR (MIR) spectral energy distributions (SEDs) to model SEDs.  The procedure described below is similar to that of \citet{ro07} and \citet{po09}.

\subsection{Data Compilation \label{dcsec}}
We started with the point source catalogs (PSCs) from the \sst\ GLIMPSE I survey \citep{be03}. GLIMPSE I (hereafter, GLIMPSE) covered the Galactic plane ($10\degr\ < l < 65\degr$, $\mid b \mid < 1\degr$) with 1\farcs2 pixel resolution in four bands, 3.6, 4.5, 5.8, and 8.0 \micron.  The GLIMPSE PSCs also include matched $JHK_{S}$ photometry from the Two Micron All Sky Survey (2MASS) \citep{sk06}. We supplemented the 2MASS near-IR photometry with point source catalogs from the latest data release (DR7) of UKIDSS. UKIDSS has a finer angular resolution (0\farcs8 FWHM, 0\farcs2 pixels at K) compared to 2MASS (1\arcsec\ FWHM, 2\arcsec\ pixels), as well as greater depth. The median 5$\sigma$ limiting magnitudes for the UKIDSS Galactic plane survey (GPS) are $J = 19.77$, $H = 19.00$, and $K = 18.05$ \citep{lu08}, while 2MASS is limited to $J = 15.8$, $H = 15.1$, and $K_S = 14.3$ (S/N = 10; \citealt[][]{sk06}).In addition, we performed aperture photometry on 24 \micron\ mosaic images (1\farcs2 pixels) from the \spit\ MIPSGAL survey \citep{ca09}. The custom code calculated 24 \micron\ fluxes and errors for the input source list using the location of GLIMPSE PSC detections and standard DAOPHOT-type routines. The relevant photometry parameters we adopted were an aperture of 7\arcsec, an annulus of 7 -- 13 \arcsec, an aperture correction of 2.05, and a zero-point flux of 7.17 Jy. If the extracted flux had a signal-to-noise ratio (S/N) less than five, then an upper limit, defined as five times the flux error, was set instead. 

UKIDSS matches to GLIMPSE sources were found by summing the astrometric errors (0\farcs09 and 0\farcs3 for UKIDSS and GLIMPSE) in quadrature, which yields a matching radius of 0\farcs313. If a UKIDSS match is found, then the UKIDSS photometry is used preferentially over 2MASS unless the magnitudes approach the saturation limit. The UKIDSS GPS has $J$ = 13.25, $H$ = 12.75, and $K$ = 12.0 as saturation limits \citep{lu08}, and we set limits 0.5 magnitudes fainter than the quoted values. If a 2MASS magnitude is brighter than $J$ = 13.75, $H$ = 13.25, and $K_{s}$ = 12.5, then the 2MASS magnitude is used for that band. 

The higher resolution and depth of UKIDSS images occasionally reveals single 2MASS/GLIMPSE objects to be multiple sources and increases the chance of a false match when trying to merge the data. If multiple UKIDSS objects are matched to a single GLIMPSE source, then the same GLIMPSE source is entered into the source list with different UKIDSS data points. However, for this study no multiply matched sources appear in the final YSO list.

We require detections in a minimum of three bands to be included in the source list. Sources with three detections may still show a true IR excess even if their SEDs and properties are not well constrained. We fixed the minimum photometric error in any band to be 10\% to prevent a single data point from dominating the resulting fits and to account for possible errors in absolute flux calibration \citep{ro07,po09}. Also, given the difficulty in extracting accurate MIPSGAL photometry, we use a minimum error of 15\% for 24 \micron\, as in \citet{po09}. We did not make any color or magnitude cuts to our sources prior to SED fitting.

\subsection{Fitting Procedure}
Source SEDs from the initial list were fit to \citet{ku93} stellar models using the routines of \citet{ro07}. We allowed the extinction to vary from \av\ = 0--35 magnitudes and picked $\chi^{2}$/n$_{data}~\leq~2$ to be the threshold for a good stellar fit. Sources having SEDs well-fit by reddened stellar photosphere were removed. GLIMPSE sources with multiple UKIDSS matches were removed if one of them was well-fit to the stellar models. We also eliminated sources where the IR excess appeared to be caused by a single band with large photometric errors.

Sources that could not be matched to reddened stellar photosphere were then fit against a grid of 200,000 YSO model SEDs \citep{ro06} that were computed using radiation transfer code presented in \citet{wh03}. YSO fitting used the same toolkit as stellar fitting \citep{ro07}. We chose model aperture sizes of 3\arcsec\ for 2MASS and GLIMPSE bands, 1\farcs5 for UKIDSS, and 7\arcsec\ for 24 \micron\ and set the distance range to 2.4 to 3.0 kpc. The aperture and distance range eliminates models with envelopes that would be extended and/or partially resolved at the estimated distance to the region. The interstellar extinction was allowed to vary from \av\ = 5--40 magnitudes, where the range was estimated from the minimum and maximum extinction values estimated from the \her\ maps. After fitting, we verified the IR excess by-eye in the images and SED plots. In total, we find 162 YSOs candidates in the \sfr\ field. Table~\ref{ysocolors} lists the YSO ID number we assigned, the GLIMPSE catalog ID, Galactic longitude and latitude, and the observed magnitudes. The $J$, $H$, and $K$ columns represent both 2MASS and UKIDSS magntidues according the selection rules from Section~\ref{dcsec}.

\subsection{Physical YSO Properties}
The YSO models are calculated over a range of properties including, but not limited to: the total bolometric luminosity, orientation, mass of the central star, disk mass, and envelope accretion rate. Since each SED can generally be fit to many YSO models equally well, it is necessary to calculate the average YSO properties for a range of well-fit models. We chose models where ($\chi^{2} - \chi_{best}^{2}$)/n$_{data}~<~1$ to be acceptable fits. The weighted average was calculated by:
\begin{equation}
P_{i}~=~e^{-\frac{(\chi_{i}^{2} - \chi_{best}^{2})}{2}} , 
\end{equation}
\begin{equation}
A_X~=~\frac{\sum X_{i}P_{i}}{\sum P_{i}} ,
\end{equation}
\begin{equation}
\sigma^{2}_{X}~=~\frac{\sum (X_{i} - A_x)^{2}P_{i}}{\sum P_{i}}~\times~\frac{N}{N - 1} , 
\end{equation}
where P$_{i}$ is the probability relative to the best fit model, X is the quantity being averaged, A$_{X}$ is the weighted average, N is the number of acceptable model fits, and $\sigma^{2}_{X}$ is the variance in the weighted average. Table~\ref{ysoparams} gives values a few important YSO properties. Column 1 is the YSO index; column 2 is an number where each digit represents a magnitude (column) from Table~\ref{ysocolors}. A '0' indicates that the band was not used in the fit, a '1' indicates the band was used, and '3' means the band was used as an upper limit, not a true detection. Column 3 is the number of acceptable fits, and column 4 is the $\chi^{2}$ of the best fit model. Columns 5 through 12 give the best fit and average values of the stellar mass, disk mass, envelope accretion rate, and integer stage. When we refer to YSO stages for individual sources, we use the stage given in column 12. Columns 13 and 14 are the average fractional stage (F$_{stage}$) and dispersion in the acceptable fitted stages ($\sigma_{stage}$). The stage classification is based off the criteria givein by \citet{ro06}. They assign stage 0/I (hereafter, I) for heavily embedded sources with an envelope accretion rate $\dot{M}/M_{*}~>~10^{-6}$~yr$^{-1}$. Stage II YSOs have $\dot{M}/M_{*}~<~10^{-6}~$yr$^{-1}$ and $M_{disk}/M_{*}~>~10^{-6}$. These objects have accreted most of their mass, but are young enough to still have a thick disk. Thin disk, stage III, YSOs have only a moderate IR excess and are characterized by $\dot{M}/M_{*}~<~10^{-6}$~yr$^{-1}$ and $M_{disk}/M_{*}~<~10^{-6}$. Since YSO evolution is a continuous process, we have calculated the average fractional stage (F$_{stage}$) and uncertainty ($\sigma_{stage}$) in addition to the integer stage. F$_{stage}$ and $\sigma_{stage}$ give a representation of which models fit the object and how well constrained they are. In some cases $\sigma_{stage}~=~0$; this does not imply a perfect fit. Rather it means that all of the well-fit models for that object give the same stage. This typically occurs when there are a limited number of acceptable fits. We calculate the integer stage by requiring that at least two thirds of the total weight (probability) be matched with a single stage:
\begin{equation}
\frac{\sum P_{i}(stage)}{\sum P_{i}(total)}~>~0.67~~~\mathrm{,}
\end{equation}
so, if the combined weight of all stage I fits for a given object is greater than 0.67, that object is classified as stage I. If no stage has a combined weight greater than 0.67, then that object is classified as ambiguous. Using this method, we find 54 stage I, 70 stage II, 6 stage III, and 32 ambiguous sources. Figure~\ref{stages} shows an example SED for each of the YSO stages. The upper-left plot shows the SED for Y133 (stage I), the upper-right shows Y112 (stage II), the lower-left shows Y157 (stage III), and the lower-right shows an Y30 (ambiguous). The solid black curve is the full SED best-fit YSO model, the grey curves other well-fit model SEDs with ($\chi^{2} - \chi_{best}^{2}$)/n$_{data}~<~1$, and the long-dashed curve is the SED for the underlying star if there were no IR excess. There are a wide range of well-fit SEDs for the ambiguous sources, which explains the difficulty in obtaining a stage I or stage ii classification. The plots clearly show the importance of far-IR photometry in constraining YSO stages.

It is important to note that while the model SED fitter can give estimates of various YSO physical parameters, the estimates become more uncertain as the number of data points and wavelength coverage decreases. \citet{de12} showed that the SED fitter may significantly overestimate the mass of heavily embedded YSOs with large envelopes, such as the stage I YSOs present in the IRDCs. It is also possible that there undetected multiple sources, so luminosities and mass may be overestimated in these cases. Thus, the physical parameters given in Table~\ref{ysoparams} should be considered carefully.

\subsection{\sfr\ YSOs}
Color-color and color-magnitude diagrams (CCDs and CMDs) are also useful for identifying and classifying YSOs, and there are several different schemes employed \citep{al04,rh06,gu08}. Figure~\ref{colors} plots mid-IR CCDs and CMDs of YSOs identified through their full SEDs. YSOs are shown as plusses for stage I, triangles for stage II, squares for stage III, and circles for ambiguous YSOs. Grey symbols indicate that the only have an upper limit at 24 \micron. The arrows show an extinction of \av\ = 15 magnitudes \citep{in05} The colors and magnitudes were chosen to provide the greatest separation between YSOs and main-sequence (MS) stars and between different YSO stages. The $[5.8] - [8.0]$ color shows excellent separation between YSOs and the MS locus in both the CMD and CCD. The YSO stages do not separate well, although the stage I objects are more red in $[3.6] - [4.5]$ than stage II or ambiguous YSOs. The $[3.6] - [4.5]$ versus $[8.0] - [24]$ CCD is much better at separating stage I YSOs from stage II YSOs. It is interesting to note that in each of the four diagrams the ambiguous objects tend to fill the gaps in between stage I and stage II YSOs. This may indicate that they are in a transition phase between classical stage I and stage II. This is also evident in the average properties in Table~\ref{ysoparams}. Four YSOs stand out with  $[5.8] - [8.0] < 0.25$ (Y36, Y50, Y117, and Y130). Y36 has a weak [5.8] detection and Y117 has a weak 8 \micron\ detection, which is the likely cause of the blue color. Y50 shows a broad dip in flux near 9 \micron, which may be a result of self-absorption from viewing an edge-on disk \citep{wh03}. The fits to Y130 show a more narrow dip in flux, which may be caused by silicates at 9.8 \micron\ \citep{wh03}.

There are several possible contaminants in any YSO sample, the most common of which are asymptotic giant branch (AGB) stars \citep{ro08}. AGB stars represent an advanced stage in the evolution of low- and intermediate-mass stars \citep{he05}. They are also intrinsically cool and are prone to mass loss resulting in significant dust formation \citep{gr06}, and so they radiate strongly in the IR with YSO-like near- and mid-IR colors \citep{wh08}. \citet{ro08} published a catalog of intrinsically red sources for GLIMPSE I/II, which includes preliminary classifications for YSOs, AGB stars, and unresolved planetary nebulae and galaxies. One of the YSOs in the initial list was listed as an AGB star in the catalog and was removed from the final YSO list. \citet{wh08} found that most AGB stars could be removed by eliminating sources with [8.0] - [24]$ < $2.2. There are 131 YSOs with both 8 and 24 \micron\ detections and only three (2\%) fall below this color criterion. In addition, AGB stars are also likely to outnumber YSOs when considering sources with [4.5]$ < $ 8 \citep{po09}; only one YSO is brighter than this threshold as the brightest sources typically presented SEDs that were removed earlier in the YSO selection process. 

In order to estimate possible contamination of our list by unassociated field YSOs we performed the same analysis on a nearby, quiescent field region. It was a circular region equal in area to the target region with a radius of 12\farcs3 centered on $l$ = 38\fdg4490 and $b$ = -0\fdg4548. For \sfr\ we found 339 sources poorly fit to reddened stellar photospheres with 162 being bona fide YSOs. In the field region, we find 146 poorly fit sources with just eight candidate YSOs. The \sfr\ region contains 176 contaminating sources (not MS stars or YSOs) and the field region contains a similar number, 138. The 28\%\ difference is most likely due to natural variation in the field star density, but a fraction of the extra sources might be YSOs that were removed due to weak IR excess or unreliable photometry. The ratio of field YSOs to SFR YSOs is quite small, 5\%, so we can assume that most of the YSOs in Tables~\ref{ysocolors} and \ref{ysoparams} are associated with the SFR itself. We also find that there are no AGB stars in the field as identified in the \citet{ro08} catalog while there was just one in the SFR, so we also conclude that the AGB contamination towards \sfr\ is small. \citet{po09} estimate $<20\%$ contamination in their YSO list for M17, however, our region is covers a much smaller area, is further from the Galactic center ($l\sim15\degr$ for M17), and likely has a less complicated line of sight. Therefore, we may expect to have a lower contamination fraction.

Figure~\ref{ysofig} is a three-color image of \sfr\ with UKIDSS K, 8 \micron, and 24 \micron\ in blue, green, and red, respectively. Red pluses, green triangles and cyan circles indicate stage I, stage II, and ambiguous YSOs. White contours delineate the gas surface density levels of 0.01(1.75$^{i}$), equaling 0.01, 0.0175, 0.0306, 0.0536, 0.0938, and 0.1641 \gcm. The YSOs tend to cluster near the IRDCs, with several stage I sources surrounding MCM~15. Most of the YSOs around N~74 are stage II or ambiguous, which indicates that it is older than MCM~15 and, likely, older than N~75 as well. The high level of diffuse 8 and 24 \micron\ emission coincident with MCM~15, which may be obscuring YSOs and causing a higher level of incompleteness. Given the apparent young age and signs of recent SF in this region, we might expect to find more YSOs coincident with the bright bubble rims, but this is not seen. It is possible that YSOs have not yet, or will not be, formed in the bubbles shells, or they are simply hidden amid the high IR background.

\subsection{Comparison to Other Regions}
The final YSO count, 54 stage I, 70 stage II, 6 stage III, and 32 ambiguous, is quite different than those in other known SFRs. The observed ratio of stage I YSOs to stage II of 0.76 is quite high compared with other reported values of 0.55, 0.4, 0.38, and 0.01 (\citealt[][]{ka09}, \citealt[][]{gu08}, \citealt[][]{po09}, \citealt[][]{ko08}; respectively). Each of the regions studied are very different in terms of morphology, sky coverage, and YSO identification methods, which may be responsible for some of the differences. It is also possible that this represents a possible age difference with higher stage I/II ratios representing younger regions. \citet{po09} studied M17 and \citet{ko08} studied the W5 \hii\ region, both of which have distances of $\sim2$ kpc, while the main clusters have ages of 3 Myr (M17) and 5 Myr (W5) based on the maximum ages of the most massive stars. These ages, and the 1--2 Myr ages for N~74 and 75, are significantly longer than the average lifetime of stage I YSOs (0.1--0.2 Myr; \citealt[][]{ad00}). Both M17 and W5 show evidence for multiple epochs of SF \citep{po09,ko08}, where the initial populations may have evolved beyond the embedded stage I phase lowering the overall stage I/II ratio. However, the \sfr\ SFR may still be in the first generation of SF and thus have a younger YSO population. The large fraction of YSOs coincident with IRDCs, which are sites of current SF, is also suggestive of young stars. The lack of stage III objects, and stage II to some extent, may be explained by the difficulties in detecting weak IR excesses, especially in regions of high and variable extinction. A stage III YSO SED may show little-to-no excess and appear stellar shortward of 10 \micron, and these may be removed by the initial stellar SED fitting. Stage II YSOs can show excesses at shorter wavelengths, but extinction may hinder their detection. 

\section{Breaking Down \sfr \label{regionsec}}
\subsection{Field Star Identification \label{fieldsec}}
The location of \sfr\ in the heart of the Galactic Plane makes it difficult to assess the associated stellar content, in part because of extinction that is high and highly variable. Also, field stars likely outnumber associated MS and pre-main sequence (PMS) stars and must be removed in order to recover likely cluster stars. In order to look for stellar clusters we follow the prescription of \citet{ma10} to evaluate and remove the field star contamination. 

The field star estimation tool compares a target region to a field region using a UKIDSS $JHK$ 3-dimensional color-color-magnitude diagrams (CCMD). After investigating different color and magnitude combinations, we used the $J, J - H,$ and $J - K$ magnitudes, as in \citet{ma10}. CCMDs were constructed for both the target regions containing a putative cluster (about 10\arcsec\ radius for MCM~15, MCM~16, N~75, and N~74) and a field annulus around the target center. We made the annulus as large as possible without overlapping a neighboring bubble,  1--2\arcmin\ for the outer radius. Since the target and field areas are, in general, not equal, the number of field stars is normalized by the target-to-field area ratio, A.

To estimate the field star content, the CCMD is divided to 3D bins along the magnitude and color directions. The cluster membership probability is then determined by the ratio of stars in the target bin stars to those in the field bin, $P = (N_{t} - N_{f}\times A)/N_{t}$. So, if there are five stars in the target bin and one star in the field bin, then one star in the target bin is a ``field star" and the membership probability for each target star is 0.8. In cases where the number of field stars in a cell is equal or greater than the number of target stars in the same CCMD cell, membership probabilities are zero.

In order to mitigate sensitivity to the choice of bin center and bin size, we used three different cell sizes for each color and magnitude. Bin widths were 0.15, 0.30, and 0.45 mag for $J - H$ and $J - K$, and 0.5, 0.75, and 1.0 mag for $J$. For each different bin size, we shift the cell centers in each dimension by one-third of the cell width in each direction. There are a total of six parameters (three width dimensions and three shift dimensions) each with three different values, this yields $3^{6}=729$ different grid permutations. For each permutation, the stars in the target cell are assigned a membership probability described above. They are averaged over all iterations to arrive at a final membership probability for each star in the target region. We tested our code on cluster LK10 and found that we could recover the morphology of the cluster CMD shown in Figure~5 of \citet{bo09}.

\subsection{MCM~15}
Figure~\ref{cl15} shows a three-color UKIDSS $JHK$ image of MCM~15 (left panel). White contours outline MIPS 24 \micron\ emission, and the magenta circle denotes the aperture used as the target region for finding clusters. YSOs are shown as red plusses for stage I, green triangles for stage II and cyan circles for ambiguous sources. In the image we see YSOs on the outskirts of the cluster, situated over regions of high gas surface density (Figure~\ref{radio}). There are only one or two YSOs detected in the central area of the cluster, however, the IR background is extremely high, as evidenced by the 24 \micron\ contours, so the data likely suffer from incompleteness. There appears to be an over density of stars, suggestive of a stellar cluster, accompanied by diffuse K-band emission, probably reflection nebulosity or emission from hot dust. There is a strong morphological similarity between the diffuse K-band and 8 \micron\ emission where PAH emission is the dominant source of luminosity (Figure~\ref{ysofig}).

Figure~\ref{cl15} (right panel) shows $J$ versus $J - K$ CMDs (top) and the $J - H$ versus $J - K$ CCDs (bottom) for the full target area (left), the field region (center), and the cleaned target (right). The cleaned CMD and CCD show only sources with a membership probability $>2/3$.  The filled diamond and asterisk mark the locations of an A0 and an O9.5V star along a ZAMS ischrone (solid line; \citealt[][]{ma08}). The dotted line curve marks a 1 Myr PMS isochrones \citep{si00}. All isochrones have been reddened by \av\ = 12, according to \citet{ca89}, and placed at the assumed distance to the molecular cloud (2.7 kpc). The solid arrow shows the magnitude and direction of \av\ = 5 magnitudes. The CCD shows an over density of stars having high membership probability near $J - K$ = 2.5--3.0, which implies approximately 12 magnitudes of extinction for a population of MS stars. The CMD shows an excess of stars between the ZAMS and PMS isochrones, which we interpret as additional evidence  of a young stellar cluster. The candidate cluster members show a significant dispersion in color, consistent with a high degree of differential reddening.  Some of the brightest cluster stars with high membership probability lie near the PMS isochrones, and these roughly correspond to the luminosity of mid-B stars. This is consistent with a recent study suggesting that MCM~15, which is coincident with IRAS 19012+0505, is an intermediate-mass star forming region that will not form stars earlier than a spectral type B3 \citep{ar10}.

\subsection{N~75} 
\begin{figure}
\includegraphics[width=18cm]{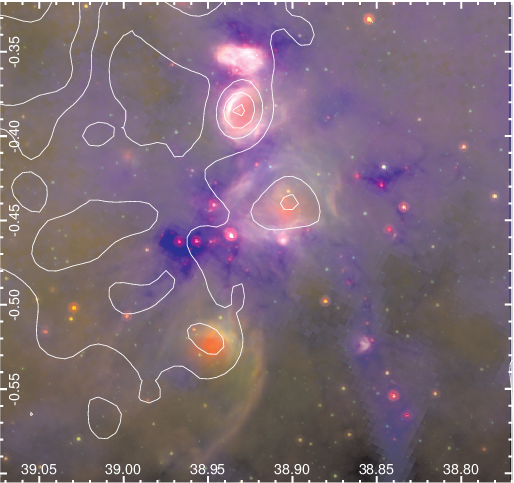}
\caption{Three-color image with \her\ 500 \micron\ in blue, IRAC 8 \micron\ in green, and MIPS 24 \micron\ \micron\ in red. White contours outline the VGPS 1.4 GHz continuum. The minimum level is the mean background temperature plus two times the standard deviation, which is 13.15 K + 2(0.43) K = 14.01 K. Subsequent contours are 4, 6, 8, 10 and 12$\sigma$ above the mean background temperature. The contours show peaks coincident with each bubble, and some emission over MCM~15 with no clear peak. \label{radio}}
\end{figure}

\begin{figure}
\includegraphics[width=18cm]{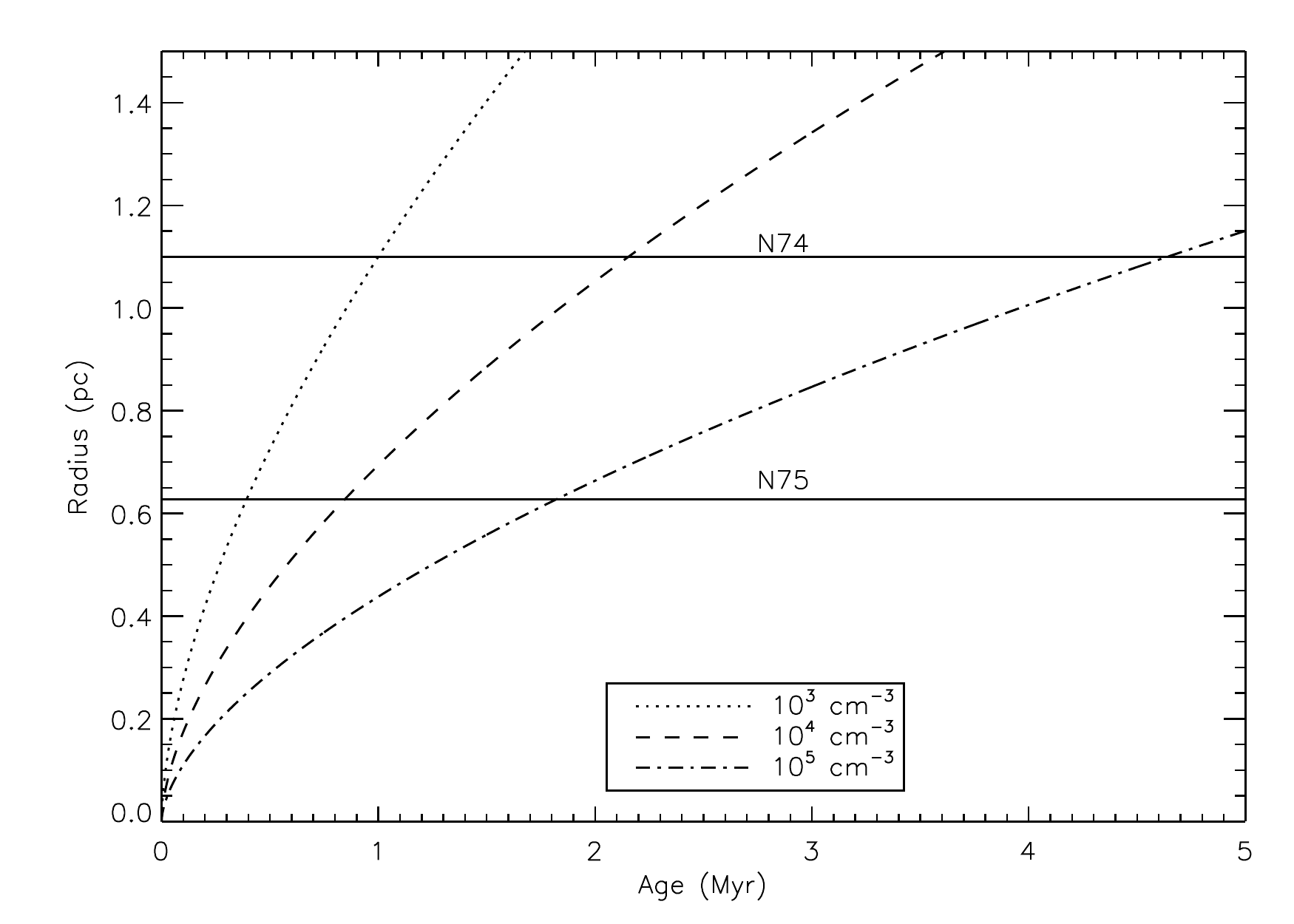}
\caption{Expansion radius versus age plot. Solid horizontal lines are the bubble radii of 0.65 pc for N~75 and 1.11 pc for N~74, while the curved lines (dotted, dashed, and dot-dashed) are the \hii\ region expansion radii for densities of 10$^{3}$, 10$^{4}$, and 10$^{5}$ \cc. \label{age}}
\end{figure}

\begin{figure}
\includegraphics[width=18cm]{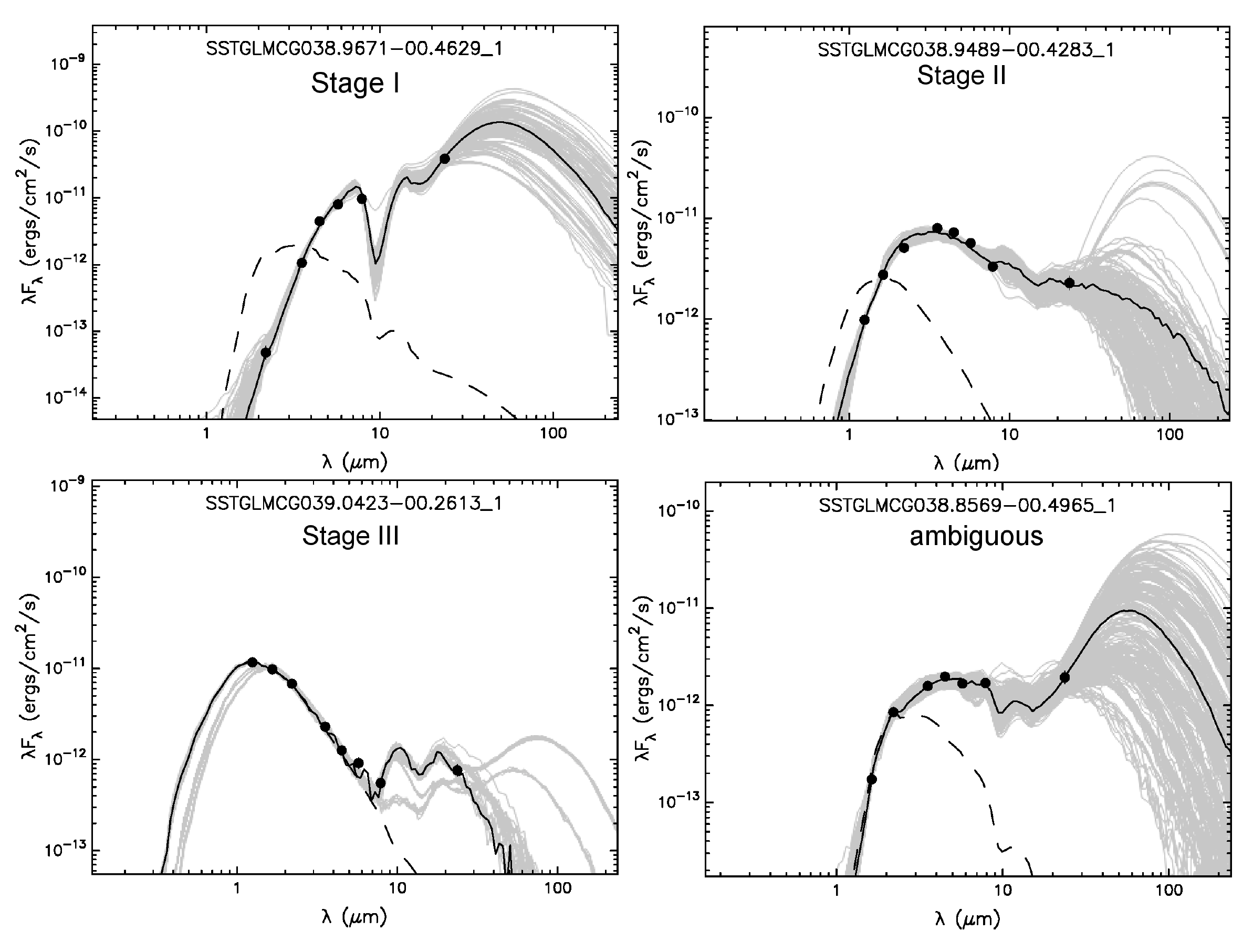}
\caption{Four-panel figure of YSO SEDs. The upper-left plot shows the SED for Y133 (stage I), the upper-right shows Y112 (stage II), the lower-left shows Y157 (stage III), and the lower-right shows an Y30 (ambiguous). The solid black curve is the best-fit YSO model, the grey curves show the range of well-fit models, and the long-dashed curve is the SED for the underlying star. \label{stages}}
\end{figure}

\begin{figure}
\includegraphics[width=18cm]{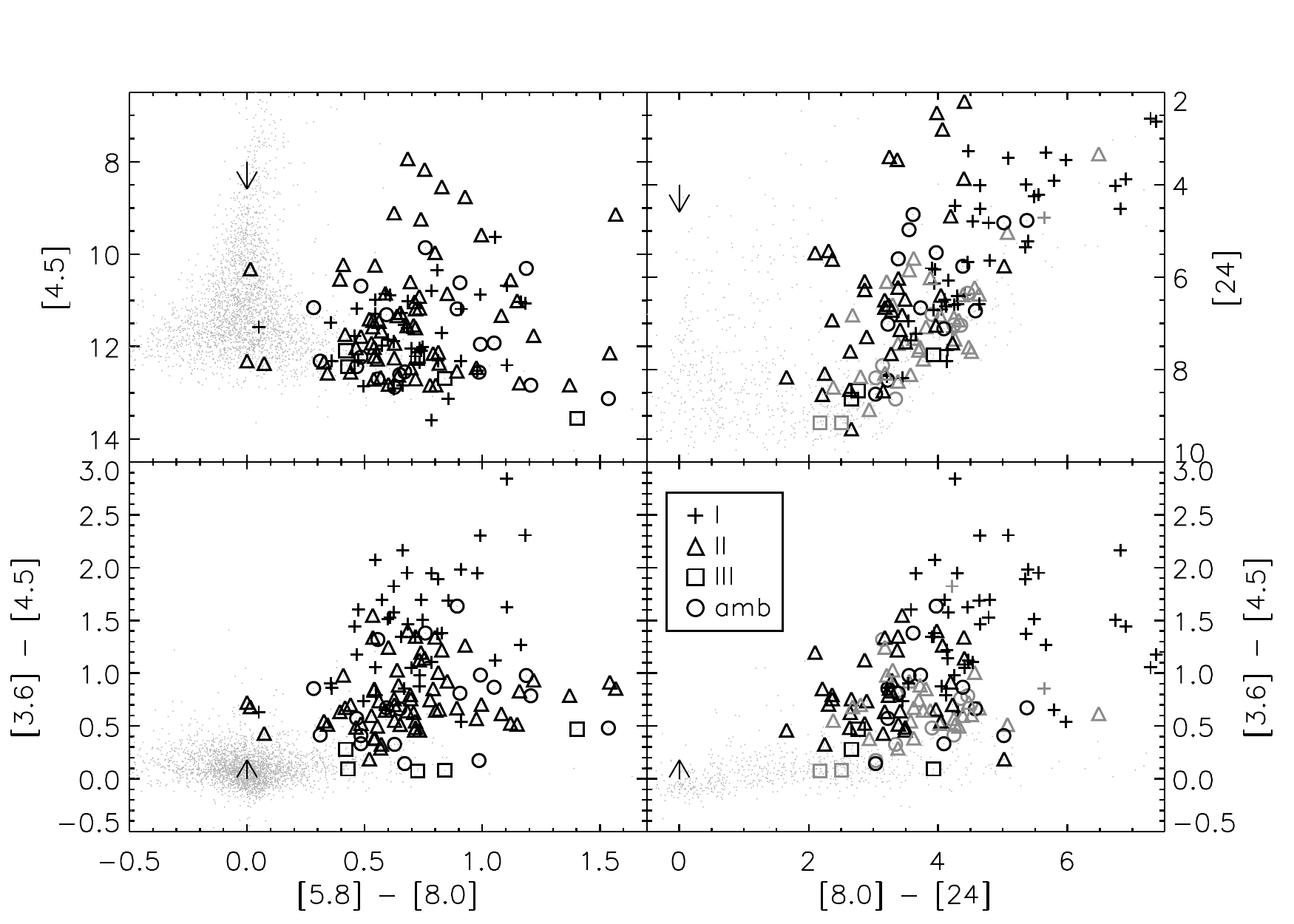}
\caption{Four-panel CCD and CMD plots using different color schemes. Stars lacking IR excess are shown as light gray dots. YSOs are labeled as plusses, triangles, squares, and circles for stage I, II, III, and ambiguous YSOs, respectively. Arrows indicate reddening vectors with \av\ = 15 magnitudes. \label{colors}}
\end{figure}

\begin{figure}
\includegraphics[width=18cm]{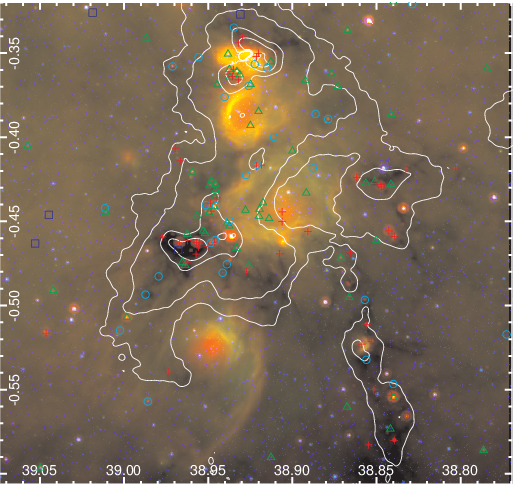}
\caption{Three-color image with UKIDSS K, IRAC 8 \micron, and MIPS 24 \micron\ in blue, green, and red. Over plotted are YSOs with stage I, II, III, and ambiguous labeled as red plusses, green triangles, blue squares, and cyan circles. White contours delineate the gas surface density levels of 0.01(1.75$^{i}$), equalling 0.01, 0.0175, 0.0306, 0.0536, 0.0938, and 0.1641 \gcm. The YSOs primarily cluster around the bubbles and IRDCs. \label{ysofig}}
\end{figure}

\begin{figure}
\centering
\subfigure{\includegraphics[width=8cm]{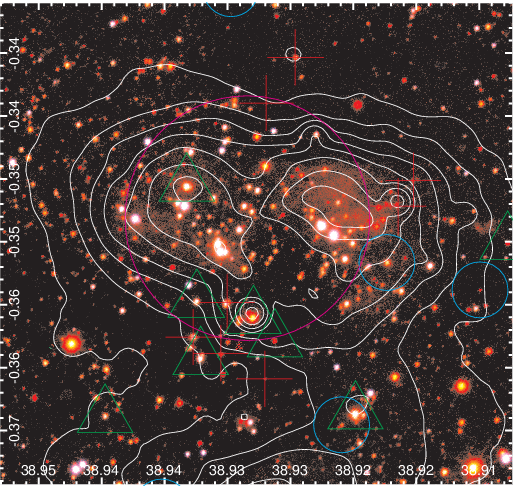}}
\subfigure{\includegraphics[width=8cm]{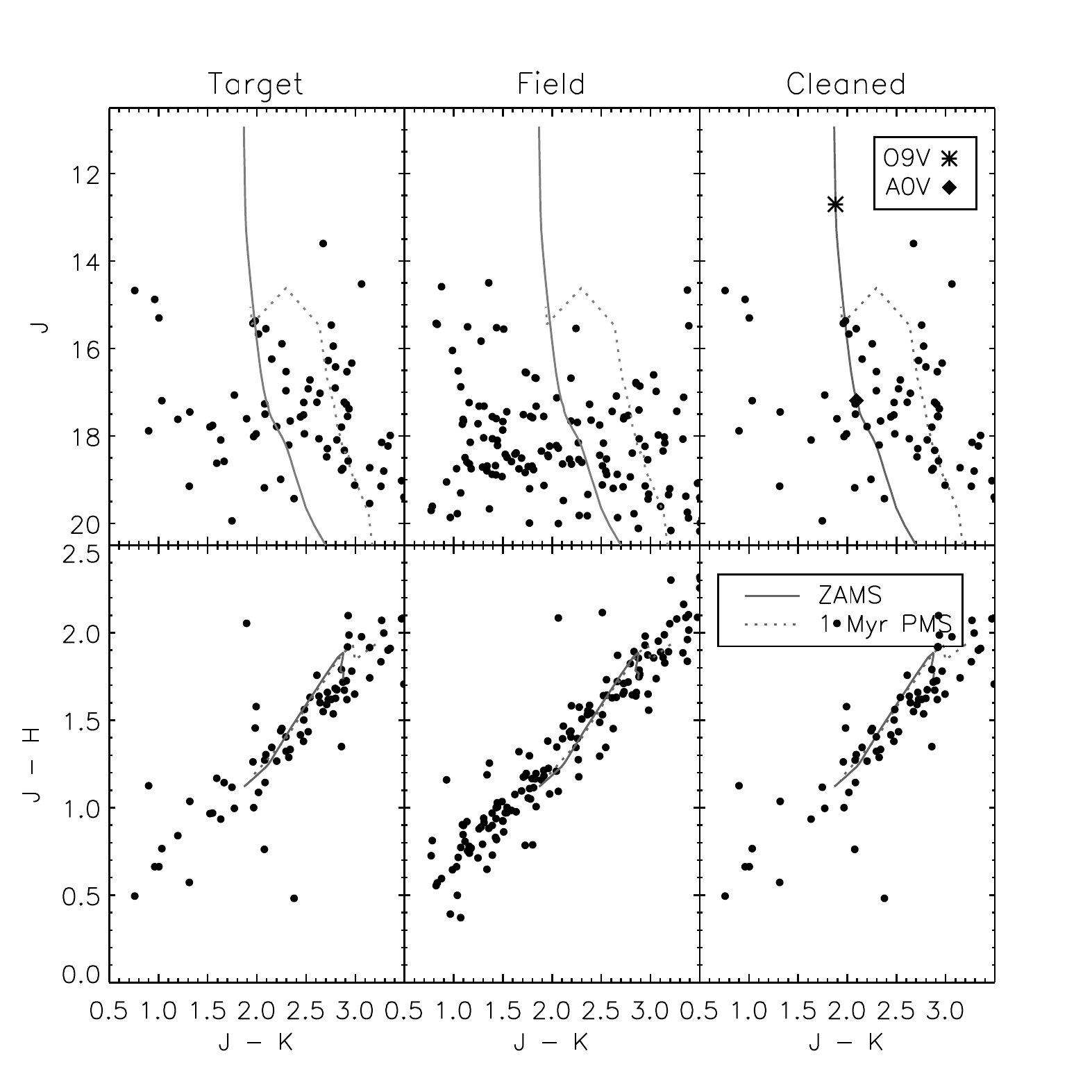}}
\caption{{\it Left}: Three-color image of MCM~15. Blue, green, and red represent UKIDSS $J$, $H$, and $K$, respectively. The magenta circle indicates the region used for cluster identification. Red plusses, green triangles, and cyan circles  represent stage I, II, and ambiguous YSOs. The contours outline MIPS 24 \micron\ emission and are spaced as: C$_{i}$ = 20($1.5^{i}$), or 20, 30, 45, 67.5, 101.3, 151.9, 227.8, 341.7, 512.6, 768.9 MJy~sr$^{-1}$. {\it Right}: CMDs (top panel) and CCDs (bottom panel) for the MCM~15 target area (left), field area (center), and after field star substraction (right).  The solid curve shows a ZAMS, and the dotted curve shows a 1 Myr PMS isochrones, both reddened by \av\ = 12 and placed at 2.7 kpc. The asterisk and filled diamond indicate the location of a O9.5 and an A0V on the ZAMS. \label{cl15}}
\end{figure}

\begin{figure}
\centering
\subfigure{\includegraphics[width=8cm]{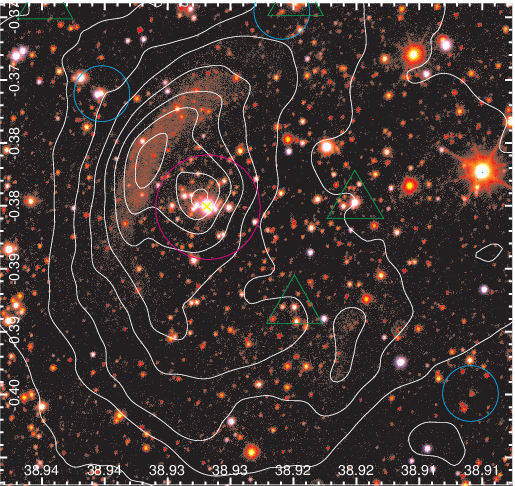}}
\subfigure{\includegraphics[width=8cm]{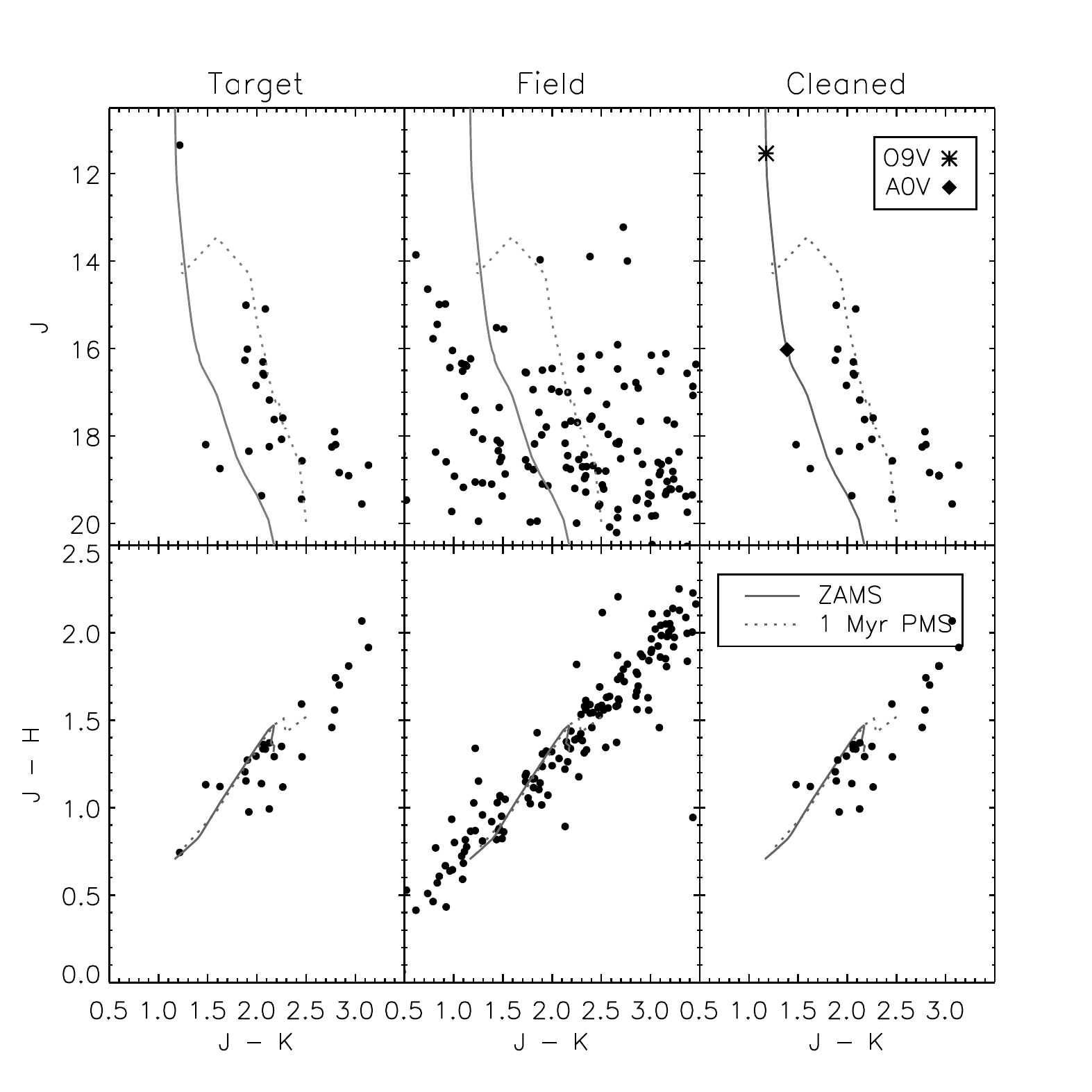}}
\caption{Same as Figure~\ref{cl15}, but for the N~75 bubble. Note the isochrone has been reddened by \av\ = 8. \label{cl75}}
\end{figure}

Figure~\ref{cl75} is a three-color image of N~75 (left panel) with color diagrams (right panel) as before. Colors and symbols are the same as Figure~\ref{cl15}. There is a possible compact cluster within the 15\arcsec\ (0.2 pc) radius magenta circle, centered on the bright star 2MASS 19034727+0509409 ($J$ = 11.35, $H$ = 10.61, $K$ = 10.13) marked by a yellow `x'. The contours showing 24 \micron\ emission exhibit a local maximum at this location, while the \her\ dust column density maps show a local minimum inside the bubble (Figure~\ref{radio}). This localized peak seen at 24 \micron\ is not visible in in the \her\ maps at 70 $\mu$m, leading us to interpret this feature as emission from very small dust grains heated locally by the star (or stars) within the compact cluster.  Interestingly, the 1.4 GHz radio continuum (Figure~\ref{radio}) also peaks at this location.

\begin{figure}
\centering
\includegraphics[width=18cm]{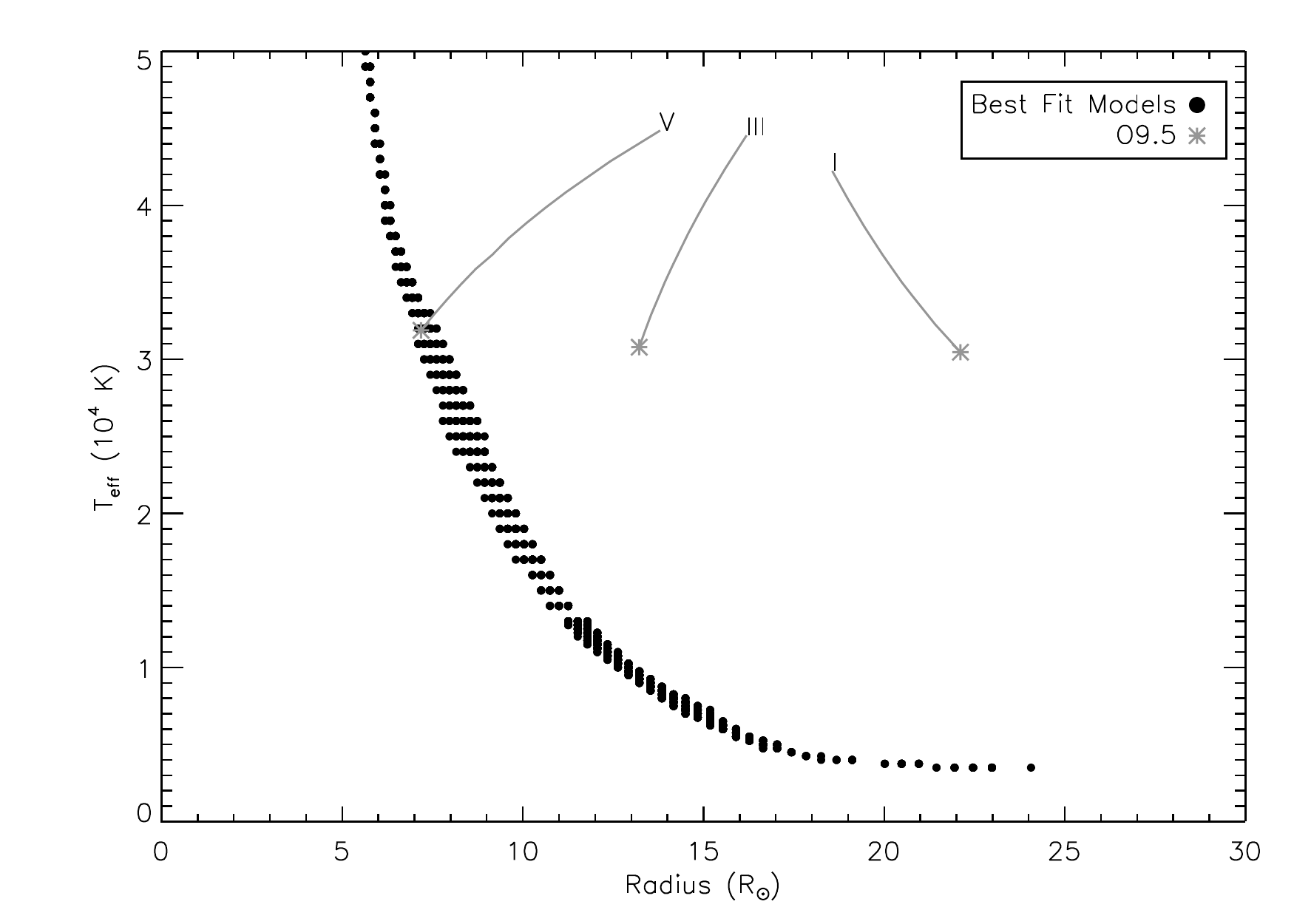}
\caption{$T_{\mathrm{eff}}-R$ relation for the candidate ionizing star in N~75. Filled black dots show the range Kurucz of stellar model that are well-fit to the star 2MASS 19034727+0509409. The gray lines indicate the $T_{\mathrm{eff}}-R$ for observed O stars, and the gray asterisks mark the location of O9.5 spectral types for the different luminosity classes. The data are consistent with 2MASS 19034727+0509409 being a late-O type star. \label{TeffN75}}
\end{figure}

The CMD (top right panel) shows a sequence of $\sim$12--15 stars that is well-matched to the PMS isochrones (2.7 kpc distance; \av\ = 8) and a small grouping of $\sim$7 extremely red stars, possibly background giants. This may indicate the presence of a young stellar cluster where the massive OB star has arrived on the MS and others are still in the disk-bearing PMS phase. The color-color diagram (lower right panel) shows a grouping of stars near $J-K$~=~2.0, consistent with a stellar cluster having similar reddening, in this case \av\ = 8 mag. The bright star, $J = 11, J - K = 1.1$ on the CMD, is consistent with a O9.5V star at 2.7 kpc with \av\ = 8 mag. Figure~\ref{TeffN75} plots the $T_{\mathrm{eff}}-R$ relation for Kurucz stellar models that are well-fit to the observed SED (black dots) at the adopted distance of 2.7 kpc. The observed $T_{\mathrm{eff}}-R$ relations for O stars from \citet{ma05} are plotted as solid gray lines and the location of O9.5V stars are shown as asterisks for the different luminosity classes. Where these two relations cross provides a good indication of the stellar spectral type \citep{wa08}, in this case a late-O type star. Its location near the symmetry axis of the MIR nebula (partially seen in Figure~\ref{cl75}) also suggests that this star may be responsible for creating the prominent $\sim$1 pc diameter IR bubble (Figure~\ref{rgb}).  Although the angular resolution of the VGPS is 1\arcmin,  the coincidence of the radio continuum peak with the 24 \micron\ peak and the candidate O9.5V star implicates this star as the source of ionizing photons responsible for generating the (presumably) free-free emission filling the bubble. The remainder of the stars in the CMD along the PMS isochrone are consistent with mid-B to early-A stars. As such, their mechanical luminosity and ionizing flux would be negligible compared to a O9.5V star. Hence, the N~75 bubble appears to surround a compact young cluster dominated by a single late-O star, but consisting primarily of intermediate-mass (and unseen lower-mass) stars. This cluster may be a descendant of an ultracompact embedded cluster \citep{al12} and a more massive version of a Herbig Ae/Be cluster \citep{te97,te99}.

\subsection{N~74}
\begin{figure}
\centering
\subfigure{\includegraphics[width=8cm]{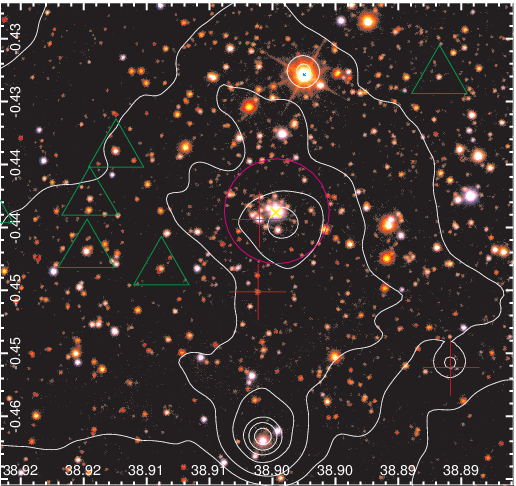}}
\subfigure{\includegraphics[width=8cm]{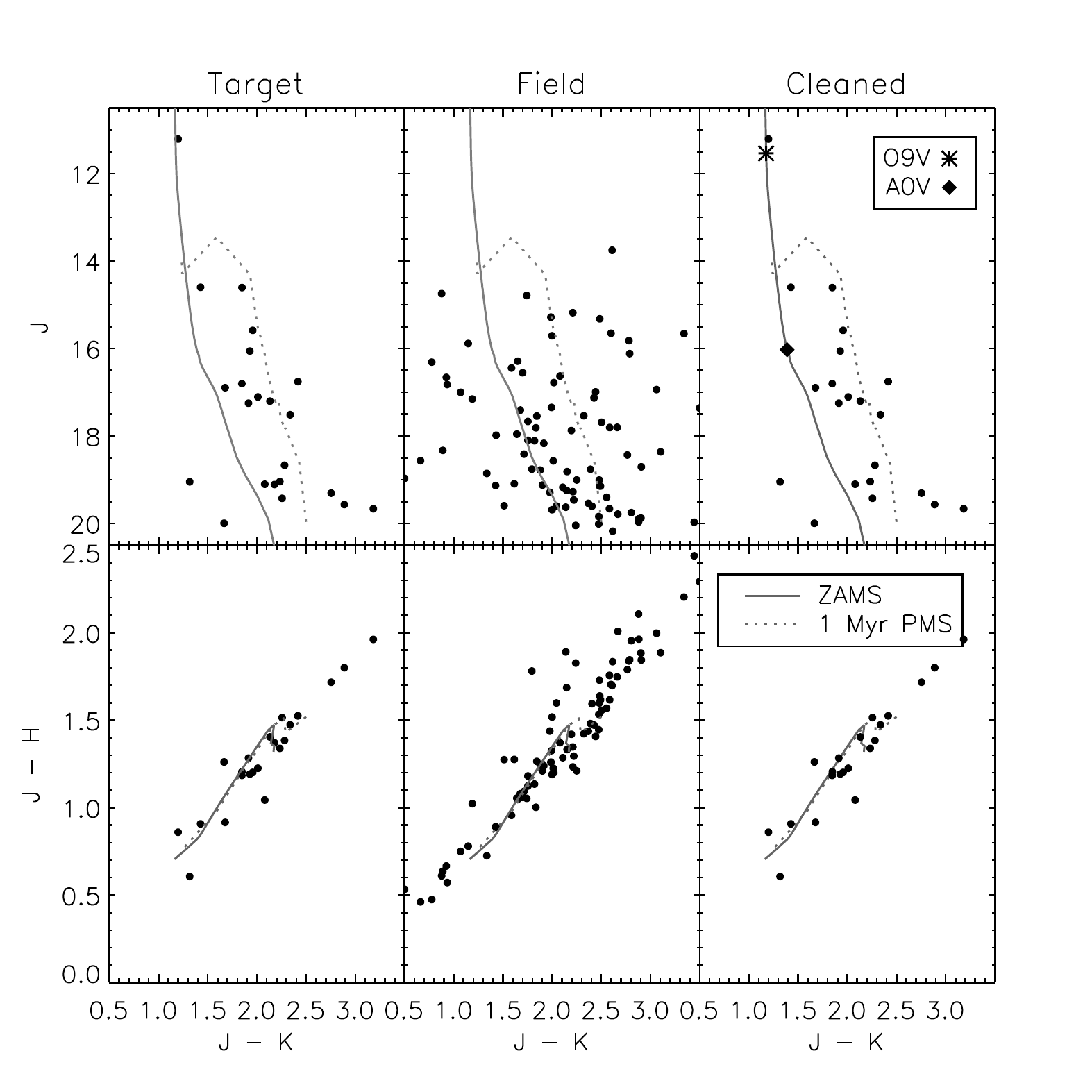}}
\caption{Same as Figure~\ref{cl15}, but for the N~74 cluster. \label{cl74}}
\end{figure}

Figure~\ref{cl74} shows the three-color JHK image (left panel) and color diagrams (right panel) for N~74; colors and symbols are the same as Figure~\ref{cl15}. The white 24 \micron\ contours peak over another potential compact cluster, anchored by the bright star 2MASS 19035684+0506370 in striking similarity to the N~75 bubble. Figure~\ref{radio} also shows a local maximum in radio continuum flux near this cluster. 

\begin{figure}
\centering
\includegraphics[width=18cm]{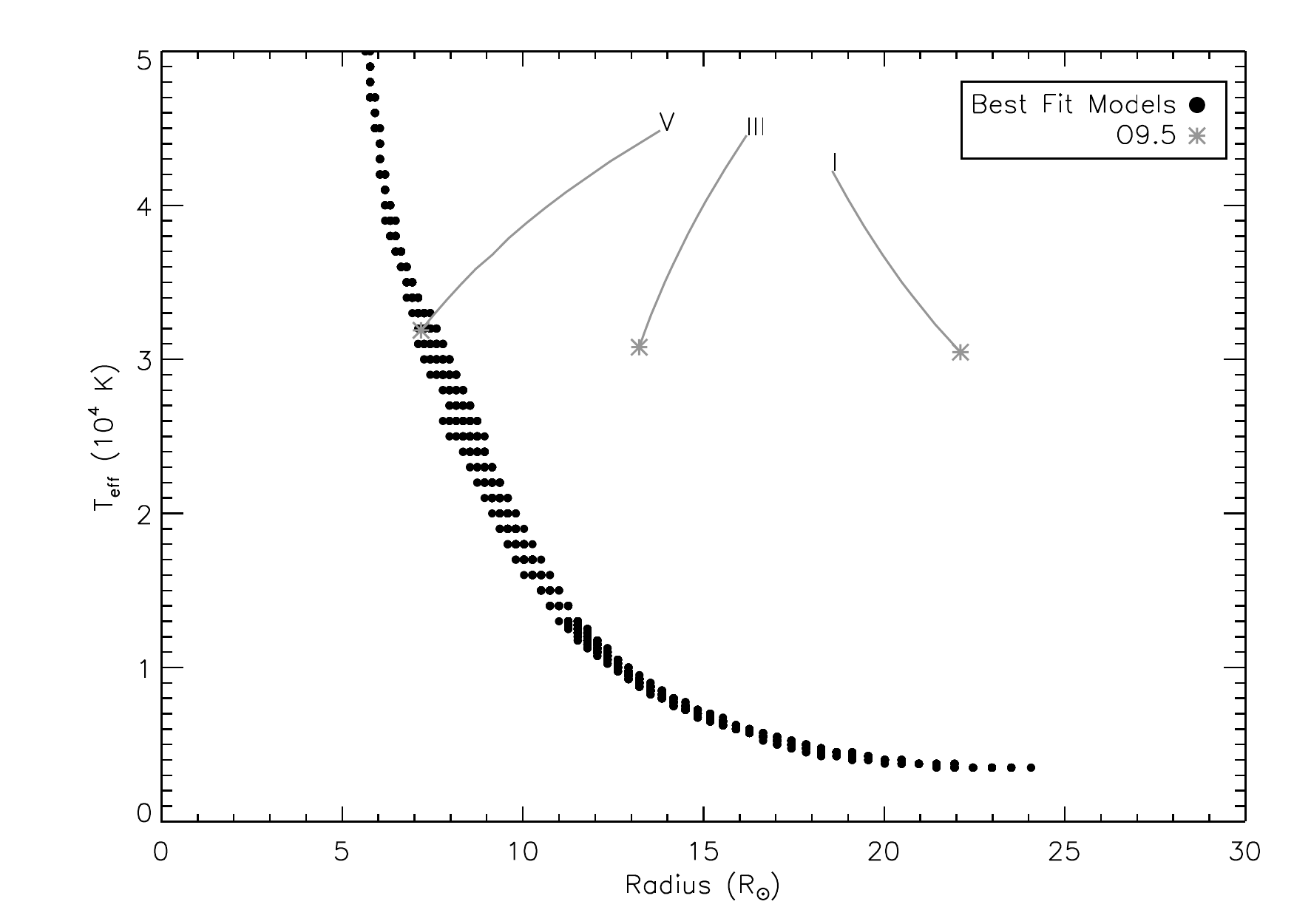}
\caption{$T_{\mathrm{eff}}-R$ relation for the candidate ionizing star in N~74, with colors and symbols as in Figure~\ref{TeffN75}.  The stellar SED fits to the bright star, 2MASS 19035684+0506370, are consistent with a late-O spectral type. \label{TeffN74}}
\end{figure}

The CMD (top right panel) shows a single bright star ($J$ = 11.21, $H$ = 10.35, $K$ = 10.01) plus a sequence of stars that is well-matched to the PMS isochrone (2.7 kpc; A$_V=$8). This is again suggestive of a stellar cluster with properties similar to that of N~75, a single O9.5V that has evolved onto the MS while its lower mass companions are still in the PMS phase. Figure~\ref{TeffN74} shows the $T_{\mathrm{eff}}-R$ relation for model fits as in Figure~\ref{TeffN75}, but for the candidate ionizing star in N~74. The data are consistent with this star having a late-O spectral type.

\subsection{MCM~16}
\begin{figure}
\centering
\subfigure{\includegraphics[width=8cm]{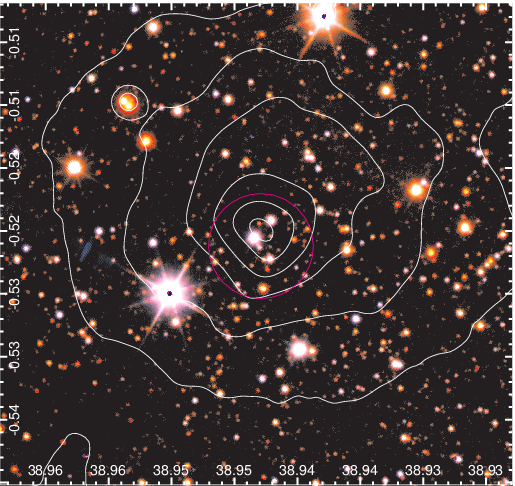}}
\subfigure{\includegraphics[width=8cm]{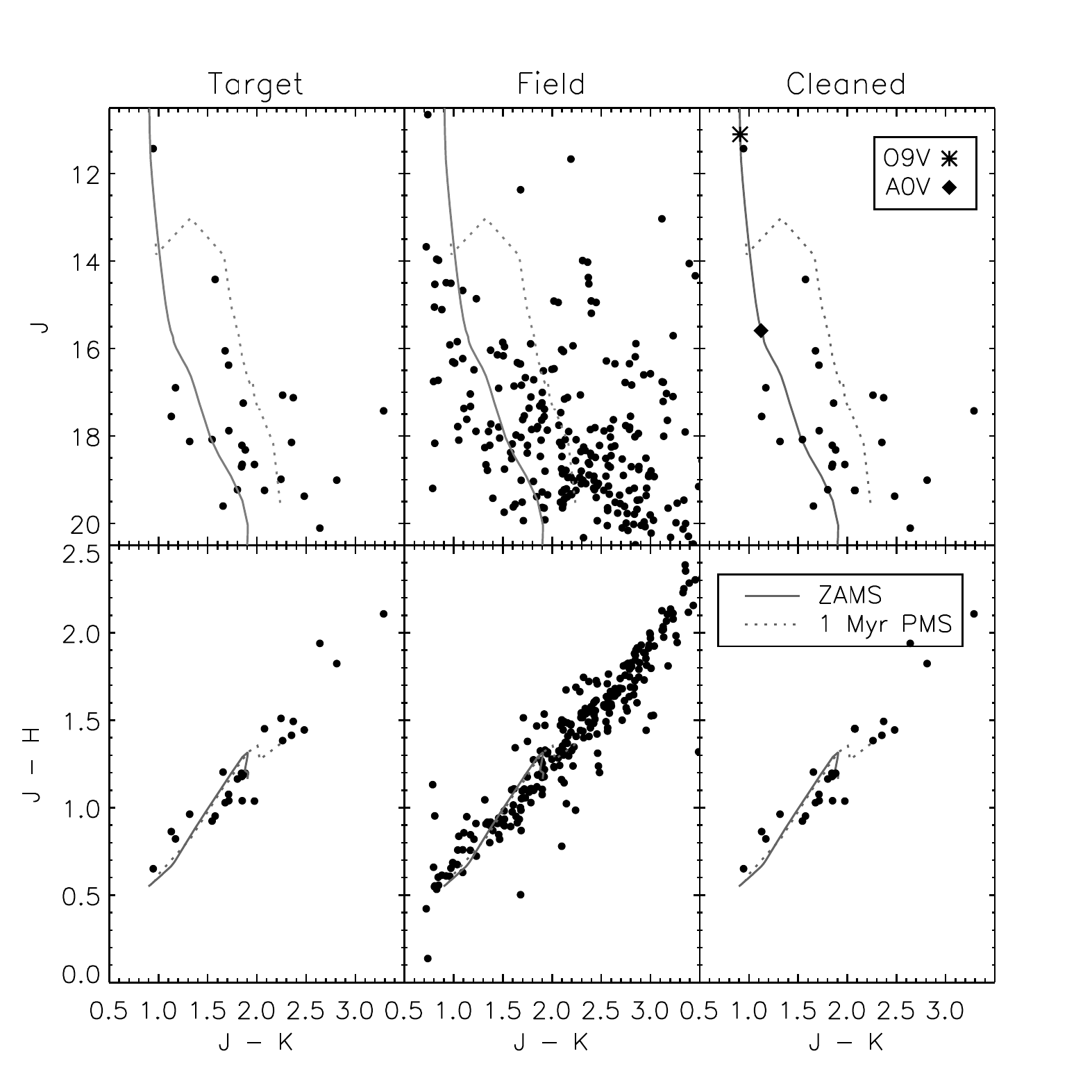}}
\caption{Same as Figure~\ref{cl15}, but for the MCM~16 cluster. \label{cl16}}
\end{figure}

Figure~\ref{cl16} shows the three-color JHK image (left panel) and color diagrams (right panel) for MCM~16; colors and symbols are the same as Figure~\ref{cl15}. The UKIDSS $JHK$ image (Figure~\ref{cl16}, left panel) again shows peaks of 24 \micron\ emission and radio continuum (Figure~\ref{radio}) centered near a bright star, 2MASS19041919+040641 ($J$ = 11.43, $H$ = 10.78, $K$ = 10.48), however, the cluster is not as compact as in N~75 and N~74. This region is further away from the main molecular complex, so there is far more stellar background contamination, as evidenced by the apparent increase in source density. This region also lacks the dense molecular gas and YSOs that are present in MCM~15, N~75, and N~74. This region is likely to be the oldest in the complex based on the extent of the mid-IR emission (Figure~\ref{rgb}). The shape of the 8 and 24 \micron\ emission does not provide a clear radius, however, we estimate the age to be 2--3 Myr based on the closest IR rim to the cluster. This more advanced age agrees with the cluster being less compact as it has passed through the gas dispersal phase and may be dissolving. 

The CMD (Figure~\ref{cl16}, top-right panel) shows a similar structure to the previous clusters, but with a lower extinction (\av\ = 6.5) and some additional scatter, likely due to the increased contamination by field stars. There is another bright star with colors consistent with an O9.5V, however, the $T_{\mathrm{eff}}-R$ relation shows this star to be inconsistent with a late-O, and it may instead be an early-B type star.

\section{Discussion \label{discsec}}
\subsection{YSO Mass Function}
\begin{figure}
\includegraphics[width=18cm]{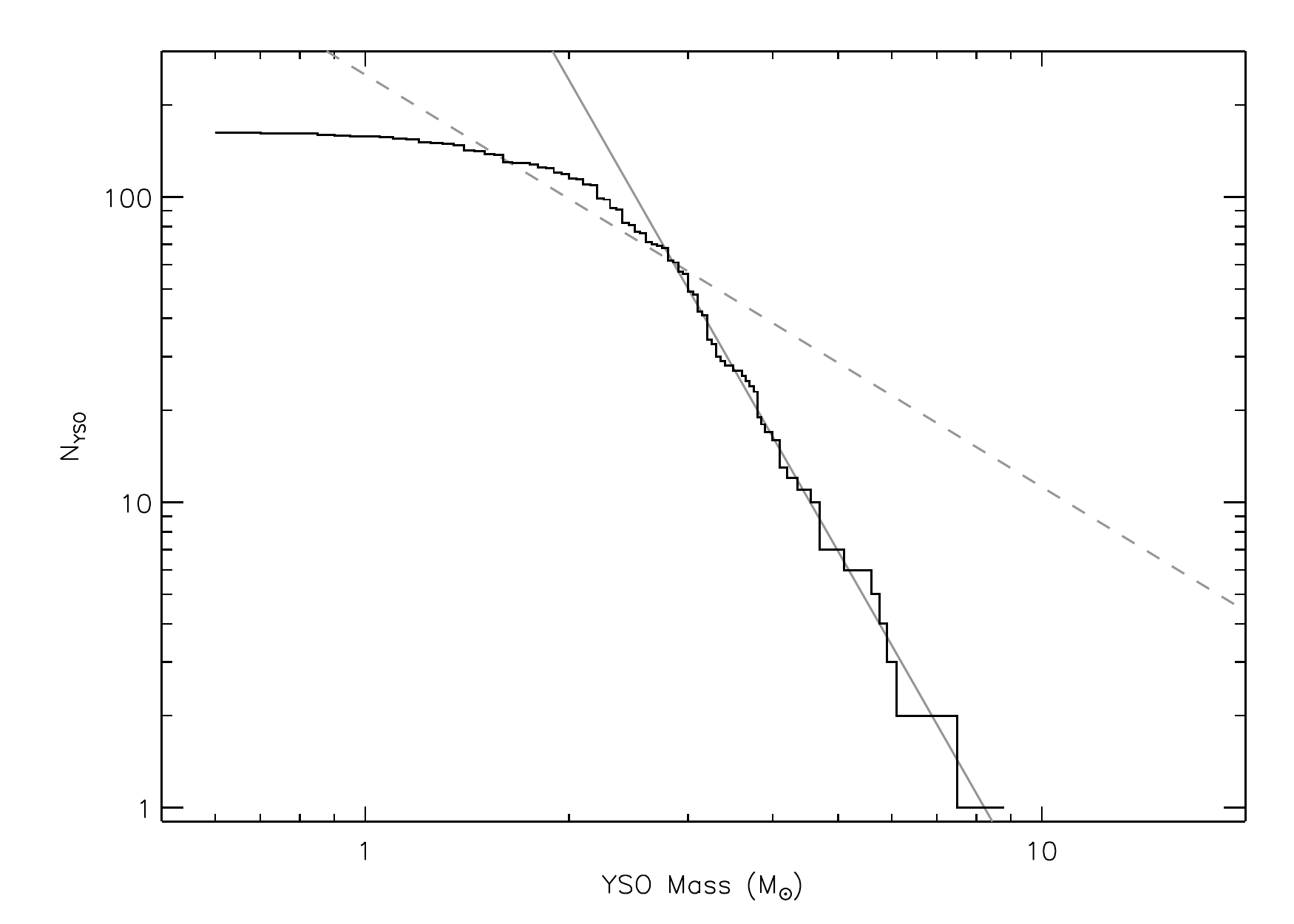}
\caption{Plot of the YSO mass function as a cumulative histogram. The number of YSOs, N$_{YSO}$, is shown as the solid black histogram, the best fit slope for YSOs $>~2.5$ \msun\ is given by the solid grey line, and a scaled Salpeter IMF is plotted as the dashed grey line. The completeness clearly begins to drop between 2.0 -- 2.5 \msun\ and becomes quite significant below $\sim2$ \msun. \label{ymf}}
\end{figure}

Figure~\ref{ymf} plots the cumulative YSO mass function (YMF). The solid black histogram is the observed YMF, the solid grey line is the best fit slope of $\Gamma_{YMF} = -3.8\pm0.1$, and the dashed grey line is the \citet{sa55} slope of $\Gamma_{S55} = -1.35$. We have chosen to use a cumulative histogram to mitigate the effects of bin size and bin location, which become important with small sample sizes (i.e., high mass bins). In addition, the slope estimation was restricted to masses $>2.5$ \msun\ to account for lack of completeness at lower masses. The resulting best fit slope is much steeper than the Salpeter IMF, but agrees quite well with the YMF found by \citet{po11} of $\Gamma_{YMF} = -3.2$ for the Carina Nebula and $\Gamma_{YMF} = -3.5$ for M17 SWex \citep{po10}. The YMF slope for our sample is likely dominated by incompleteness, especially  below $\sim2$ \msun\ where source counts drop off, but inclusion of such sources would only make the fitted slope \emph{steeper}. On the more massive side, the lack of higher mass YSOs ($>6$ \msun) may be due to stochastic sampling of the IMF or possibly that high-mass stars evolve to the MS more rapidly than low-mass stars \citep{po10}. Figures~\ref{cl75}, \ref{cl74}, and \ref{cl16} each show a likely star cluster with a massive central star. The most massive star appears to be on, or close to, the MS while the fainter stars seem to fall along the PMS tracks.

The most likely scenario for explaining the lack of massive YSOs is simply that this SFR does not present the appropriate conditions. \citet{kr08} find a minimum surface density of 1 \gcm\ for the formation of massive stars, while the highest surface density in the \her\ 500 \micron\ map is $\sim0.25$ \gcm, and values above 0.1 \gcm\ are generally limited to the IRDCs. This suggests that, at present, this SFR does not have densities high enough to form massive stars. This suggestion is supported by the classification of the MCM~15/N~75 region as an  intermediate-mass SFR \citep{ar10}.

\subsection{YSOs and Gas Surface Density}
\begin{figure}
\includegraphics[width=18cm]{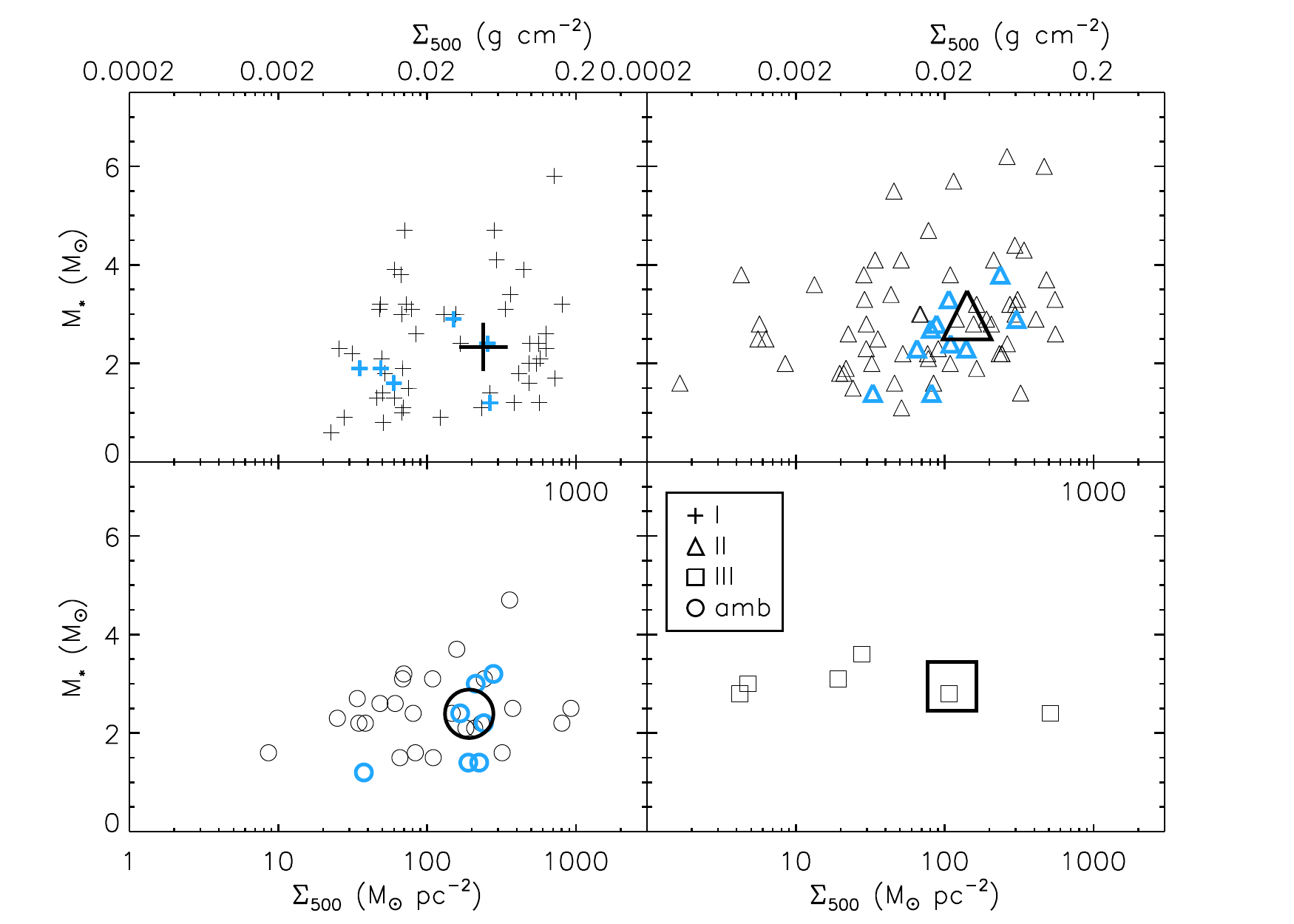}
\caption{Plot of YSO mass versus 500 \micron\ mass surface density. Stage I YSOs are plus symbols (upper left), stage II are triangles (upper right), ambiguous sources are circles (lower left), and stage III are squares (lower right). Large symbols mark the arithmetic average YSO mass and gas mass surface density for each stage. \label{nh2}}
\end{figure}

Figure~\ref{nh2} shows a plot of individual YSO mass versus gas mass surface density derived from \her\ 500 \micron\ maps. The upper left plot shows stage I YSOs, upper right shows stage II, lower left shows ambiguous, and lower right shows stage III. The heavy blue symbols are those sources within the radii of the N~75 and N~74 bubbles given in Table~\ref{params}, and extra large symbols show the arithmetic average of all the sources in each panel. Since it takes more gas to form a larger star, one might expect more massive YSOs to be embedded in higher surface density regions than less massive YSOs, however, this trend is not seen in the plots. The linear correlation coefficients are 0.22, 0.19, and 0.22 for stage I, stage II, and ambiguous YSOs, which indicates a weak degree of correlation for 54, 70, and 32 data points, respectively. There might be evidence for an upper envelope as 4 \msun\ YSOs are typically associated with a surface density higher than 50 \mpc\ (0.01 \gcm). The estimated uncertainty on an individual YSO mass ranges from 0.2--0.5 \msun\ depending on how well constrained the models are, but variations of this size appear insufficient to obscure any tight underlying correlation. However, there are likely to be other sources of scatter in the estimated masses and gas column densities. The difference in angular resolution between the GLIMPSE and \her\ 500 \micron\ beamsize (1\farcs2 vs. 36\arcsec\ FWHM) contributes uncertainty in matching a given YSO with a local gas mass surface density. The surface density maps were interpolated to the 3\farcs2 scale of the 70 \micron\ images, but this is not equivalent to sub-mm data with a small native pixel scale. If the ratio of unrelated background gas to \sfr\ molecular cloud gas varies on scales smaller than the GRS beam size, then additional pixel-to-pixel noise will be added to the surface density maps. However, we do find that, on average, stage I YSOs fall at a higher surface density ($\sim$239 \mpc) not quite twice that of stage II YSOs ($\sim$141 \mpc). The ambiguous sources fall in between these values ($\sim$191 \mpc), which may indicate that these objects are transitioning from a heavily embedded phase to a less embedded phase. While there are few stage III YSOs, it is worth noting that they fall in the lowest density regions with an average gas density of $\sim$112 \mpc.

\begin{figure}
\includegraphics[width=18cm]{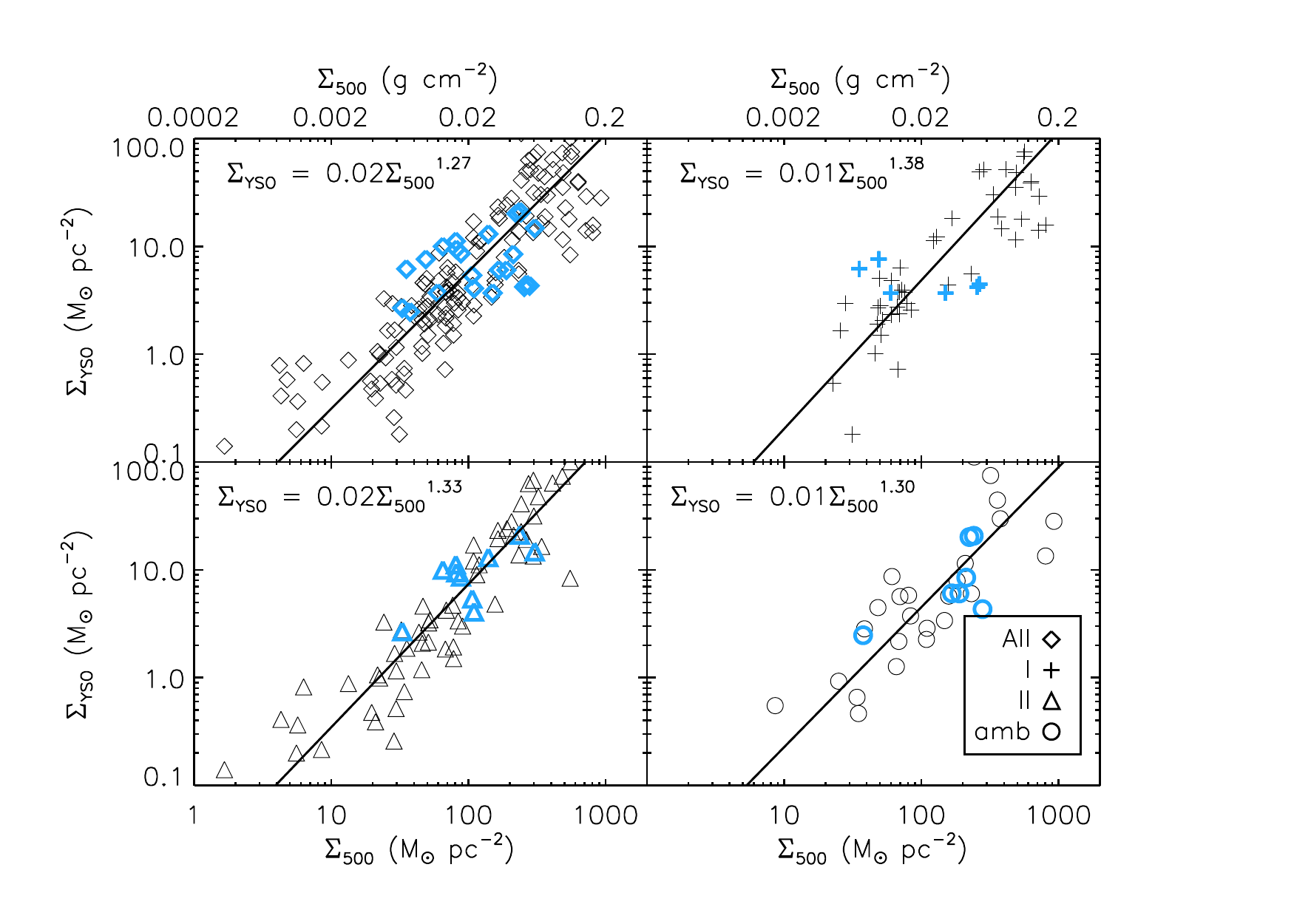}
\caption{Plot of YSO surface density versus 500 \micron\ mass surface density. The upper left shows the surface density of all YSOs, the upper right shows only stage I, lower left shows stage II and lower right shows ambiguous objects. Solid lines indicate power law fits to the data given by the equation in the upper left-hand corner of each panel. \label{surf}}
\end{figure}

Figure~\ref{surf} is a plot of YSO mass surface density against the local ISM gas mass surface density. We followed \citet{gu09} and calculated the angular radius out to the sixth nearest neighbor YSO. Then, using the adopted distance of 2.7 kpc, we converted the angular radius into a physical radius and calculated the enclosed area. The YSO mass surface density is then the sum of the masses of the N=6 YSOs divided by the physical area they cover. The upper left panel shows all of the YSOs, upper right shows stage I YSOs, lower left shows stage II, and lower right shows ambiguous sources. Heavy blue symbols are those YSOs within the N~74 and N~75 boundaries. We find excellent power-law correlations between the YSO mass densities and gas surface densities with the best fit equation given in the upper left corner of each panel. The slopes are 1.27$\pm$0.06, 1.38$\pm$0.11, 1.33$\pm$0.08, and 1.30$\pm$0.17 for all YSOs, stage I, stage II, and ambiguous sources, respectively, so within the uncertainties, the slopes are equivalent. \citet{gu11} find a wide range of slopes between 1.4 and 3.8 for eight nearby SF regions, but they did not calculate slopes for distinct evolutionary stages. They also adopt an average mass for their YSOs of 0.5 \msun, while we use the average individual mass from the model YSO fits. We also use the sixth nearest neighbors, while \citet{gu11} use up to the 11th. If the nearest neighbor count was increased from 6 to 11, then the resulting slopes all drop by about 11\%, meaning that our results are relatively insensitive to the choice of N. Another difference is that the regions in studied in \citet{gu11} have distances less than 1 kpc and thus have a greater ability to detect low-mass YSOs. Figure~\ref{ymf} shows a clear drop in YSO counts below 2 \msun, which may have significant effects on the slope of the YSO-gas density relation. If undetected YSOs have a similar spatial distribution to the current YSO list, then the slope of the density relation would remain the same but with a larger y-intercept. However, it is possible that faint (i.e., low-mass) YSOs are preferentially detected in areas with lower extinction and lower sky background.  If this is true, then a higher fraction of low-mass YSOs may remain undetected in high gas surface density regions. This type of detection bias could steepen the slope shown in Figure~\ref{surf} and bring the values more in line with those from \citet{gu11}. Interestingly, the Schmidt-Kennicutt Law \citep{ke98}, which describes the power-law relation between the areal SF rate density and the mass surface density has a power law index of 1.4 similar to the YSOs in this region. 

\subsection{Evidence for Triggered SF?}
In the Introduction, we described the basic mechanism for three theoretically defined triggered SF scenarios: CC, RDI, and RR. Another mode of triggered SF involves the sequential collapse of material along a molecular filament. Simulations by \citet{fu00} show that as an \hii\ region expands in a filament, it will cause the cloud to ``pinch" and trigger the collapse of new SF cores to either side of the original \hii\ region. If these new cores eventually form their own \hii\ regions, the process will repeat. Recent evidence has shown that SF molecular clouds tend to be filamentary in nature \citep{mo10b}, and observational studies have suggested that ``filament pinching" might trigger sequential waves of SF \citep{sa12} and work by \citet{ko12} hints at a similar sequential SF scenario. In regions with clear MS populations and known massive stars, they argue that the distribution of YSOs mirrors the pre-SF filamentary structure of the parent molecular clouds. They also suggest that massive stars and \hii\ regions enhance SF by compressing the filaments. 

The GLIMPSE 8 \micron\ and \her\ 500 \micron\ images (Figure~\ref{radio}) show that the spaces between the bubbles have dark lanes. This may indicate a swept-up shell, or, it may instead be evidence of a  pre-existing filament that extended from north of MCM~15 through the bubbles and further south. In this model one of the bubbles could have been the initial \hii\ region, which triggered the sequential collapse of the cloud. However, the estimated bubble ages are similar and it is not clear whether one would have had sufficient time to trigger the other. It may be that the bubbles formed within the same cloud but do not have a causal relationship.

In a large statistical study, \citet{th12} cross-matched the location of massive YSO (MYSO) candidates \citep{ur08} with IR bubbles \citep{ch06} and found that 22\%\ of MYSOs are coincident with bubble rims. \citet{ke12} used the more complete bubble catalog from the Milky Way Project \citep{si12} and found a similar coincidence within the bubbles. They also found that 67\%\ of MYSOs lie within 2 bubble radii. These data suggest that triggered SF occurs in the presence of bubbles and \hii\ regions. However, many bubbles are likely to reside in or near large SFRs and molecular cloud complexes. As such, bubbles are not unbiased regions distributed at random, but rather they are likely biased toward higher-than-average column density regions where stars are already expected to be forming. Figure~\ref{ysofig} shows that most of the YSOs may be associated with the IRDCs surrounding N~74 and N~75 rather than with the bubbles, even at distances exceeding one bubble radius. Most of the YSOs that lie between one and two bubble radii away from N~74 and N~75 are most likely associated with nearby IRDCs or other features and the bubbles. Therefore, extending the triggered region beyond one bubble radius increases the likelihood of including spurious YSOs that are not causally connected to the bubbles.

One method of demonstrating the occurrence of triggered SF is to establish a radial YSO stage gradient as a proxy for age. The N~74 bubble has two stage I YSOs within it compared to six stage II YSOs, while the rim is a mixture of stage I, II, and ambiguous YSOs. The bubble interior then has a stage I/II ratio of 0.2 compared to the 0.75 stage I/II ratio seen over the entirety of \sfr. This may be suggestive of an age gradient, but the limited number of YSOs around the N~74 and N~75 prohibit a statistically significant determination. We also see a dip in gas surface density (Figure~\ref{radio}) coincident with the bubble centers and a density enhancement on the periphery, which is suggestive of a swept up shell. However, the portion of the N~74 rim that shows the largest number of YSOs also abuts the IRDC, and the YSOs may have formed from the IRDC prior to the advancement of the \hii\ region. We conclude that there is no definitive evidence for an age gradient in the N~74/N~75 bubbles.  

Triggered SF via RDI and RR is characterized by a relatively short timescale ($<1$ Myr) and the presence of YSOs with trailing, wind-blown pillars pointed towards the exciting source(s). There appear to be no such structures within the entire region. It is possible that the flux of ionizing Lyman continuum radiation is not high enough to effectively compress the surrounding ISM to form stars or that not enough time has passed for YSOs to have formed through RDI. However, at the estimated bubble age of 1 Myr, RDI triggered YSOs should have had enough time to form. We conclude there is no evidence that RDI plays an important role in this SFR.  

Triggering via CC predicts that SF will occur in massive, regularly spaced clumps over Myr timescales and that the subsequent generation will include massive stars \citep{el77}. N~74 and N~75 appear to be wind-blown bubbles and the {\it Herschel} sub-mm and gas surface density maps indicate that they have been at least partially evacuated, as would be seen in a CC triggered SF scenario. However, there is no evidence for current SF within the bubble rims, and only a handful of YSOs have been detected within the bubbles, none of them massive. Also, while the gas surface density does drop within the bubbles it is not obvious that gas condensations have formed as a result of stellar feedback. The highest gas surface densities  detected near the bubble peripheries (0.05 \gcm at the upper left corner of N~75 as seen in Figure~\ref{rgb}) occur in the same azimuthal direction as the IRDCs that lie just outside the bubbles. The projection of the IRDCs and bubble rims makes it difficult to determine whether the gas is swept-up material from the bubble interior or the ambient molecular gas comprising the IRDCs, or both. We suspect  that the majority of the gas is related to the IRDCs because surface density contours match the outline of the IRDCs visible in the GLIMPSE 8 \micron\ maps. For N~74 and N~75, we conclude that there is no compelling evidence to invoke triggered SF through CC on bubble rims as a result of the stellar feedback.

\subsection{Does stellar feedback enhance or suppress SF?} 

If feedback enhances SF through triggering, one possible observable consequence is a higher star formation efficiency (SFE), i.e., a higher density of YSOs per unit gas mass in triggered SFRs than in non-triggered regions. That is, triggering may elevate the SFE by precipitating collapse of clouds that are marginally stable and otherwise would not have formed stars. Numerical simulations suggest that stellar feedback may both trigger and suppress SF simultaneously and not have a strong effect on the overall SFE \citep{da12}. Our analysis is consistent with a scenario whereby stellar feedback does not have an appreciable net effect on the total SFE of a molecular cloud. Figure~\ref{surf} does not show any significant difference in the density of YSOs on the rims or within bubbles (blue symbols) compared to those not obviously influenced by massive star feedback (black symbols; regions outside the bubbles). \emph{Although the statistical power in this sample is limited, this null result may be taken to mean that triggering either has not occurred, or that feedback does not affect the SFE and is not as significant a factor as the initial molecular cloud density.} This conclusion is in agreement with others who have suggested that triggering does not measurably affect the global SF rate \citep{el11}. Stellar feedback-induced triggering may be just one of many mechanisms to form stars among processes such as the isolated spontaneous collapse of molecular cores, SF driven by cloud-cloud collisions, or other dynamical effects. In essence, SF within a cloud does not depend on how the molecular gas reaches the critical density, only that the density is reached before the SF clumps are disrupted by outside forces \citep{el11}. Our analysis of SFR within \sfr\ appears consistent with this hypothesis.

\section{Conclusions \label{concsec}}
We have performed a multiwavelength analysis on the \sfr\ SFR, which lies along a relatively unconfused line of sight at a distance of 2.7 kpc. The complex hosts a series of IR-bright bubbles, a web of IRDCs, and a large molecular cloud. 

\begin{enumerate}
\item The maximum gas mass surface density within the region is 0.2 \gcm. This suggests that current SF will is limited to low- and intermediate-mass stars, consistent with observations of the YSOs.
\item The N~74 and N~75 bubbles are each powered by a single O9.5V star surrounded by a compact cluster. There is a similar cluster in MCM~16, but likely anchored by an early-B star.
\item We detect a total of 162 YSOs with 54 stage I, 70 stage II, 6 stage III, and 32 ambiguous classifications. The high ratio of stage I to stage II YSOs (0.76) is consistent with the idea that SF in this region has only recently begun.
\item The calculated YSO mass function slope $\alpha=-3.8$ is significantly steeper than Salpeter but consistent with other YMF studies.
\item We find only a weak correlation between individual YSO mass and the local gas mass surface density.
\item There is, however, a strong correlation between areal YSO mass surface density and gas mass surface density with a tight power-law slope of 1.3.
\end{enumerate}

Classical signposts of triggering (pillars, YSOs coincident with bubble rims \& shells, YSO age gradients within bubbles) appear to be absent in this region. We conclude that, at present, traditional mechanisms (CC, RDI) proposed as triggers for SF have likely not occurred. This may be the case because either the single O9.5V stars provide insufficient mechanical/luminous energy to initiate triggering, or because the bubbles are too young for triggering to have begun, but we cannot discriminate between these possibilities at present.  We used the bubbles to divide \sfr\ into possible causal (triggered) and non-causal (non-triggered) regions to investigate the SF properties separately. We find no evidence to suggest that stellar feedback has enhanced or supressed the local SFE within the bubbles relative to other regions of the molecular cloud. This suggests that the SF rate and SFE are sensitive mainly to the gas density, regardless how the gas is collected. Detailed studies of the molecular gas around a larger sample of \hii\ regions should be able to quantify the importance of feedback-triggered SF by determining the fraction of high-density, feedback-compressed gas surrounding \hii\ regions relative to that already present in quiescent molecular clouds. Such studies would yield a quantitative measure of the fraction of all SF possibly influenced by stellar feedback.  

\acknowledgments 
MJA is supported by NASA GSRP fellowship NNX 10-AM10H. HAK and CRK are supported by NASA through grant ADAP-NNX10AD55G. The authors would like to thank Tom Robitaille for help with the YSO fitter and Matt Povich for his guidance in classifying YSOs. Thanks to Ed Churchwell and Barb Whitney for many helpful comments. The manuscript was significantly improved by the comments of an expert referee.

\clearpage
\begin{deluxetable}{lccccccccccc}
\tablecaption{\sfr\ Region Parameters \label{params}}
\tablewidth{16.0 cm}
\tabletypesize{\scriptsize}
\tablehead{\colhead{Name}& \colhead{l}& \colhead{b}& \colhead{R}& \colhead{V}& \colhead{T}& \colhead{M$_{^{13}CO}$}& \colhead{M$_{500}$}& \colhead{Flux (20 cm)}& \colhead{log$_{10}$(N$_{Ly}$)}& \colhead{N(O9.5V)}& \colhead{t$_{6}$}\\
\colhead{}& \colhead{\degr}& \colhead{\degr}& \colhead{\arcmin}& \colhead{\kms}& \colhead{K}& \colhead{\msun}& \colhead{\msun}& \colhead{mJy}& \colhead{ph s$^{-1}$}& \colhead{}& \colhead{Myr}\\
\colhead{(1)}& \colhead{(2)}& \colhead{(3)}& \colhead{(4)}& \colhead{(5)}& \colhead{(6)}& \colhead{(7)}& \colhead{(8)}& \colhead{(9)}& \colhead{(10)}& \colhead{(11)}& \colhead{(12)}
}
\startdata
MCM~15& 38.9308& -0.3525& 0.9& 39.0&28$\pm$6& 519\err174&666\err200&\nodata&\nodata&\nodata& $<$0.5\\
N~75& 38.9262& -0.3875& 1.2& 40.0&26$\pm$3& 824\err276&402\err120&35.3$\pm$4.7&46.33& 0.03&0.4 -- 0.8\\
N~74& 38.9085& -0.4378& 1.9& 40.3&24$\pm$1& 1080\err362&699\err210&19.2$\pm$3.5&46.06& 0.02&1.0 -- 2.2\\
IRDC& 38.9672& -0.4685& 1.8& 41.8&21$\pm2$& 1941\err651&1386\err418&\nodata&\nodata &\nodata &\nodata\\
MCM~16& 38.9518& -0.5352& 1.3& 40.7&24$\pm$2& 365\err122&84\err25 &11.0$\pm$2.8&45.82& 0.01& $>$2\\
Whole& 38.9202& -0.4411& 9.1& 41.0&21$\pm$2& 18635\err6250&9221\err2766&\nodata&\nodata &\nodata &\nodata\\
\enddata
\end{deluxetable}

\clearpage
\begin{deluxetable}{llcccccccccc}
\tablecaption{List of Candidate YSOs \label{ysocolors}}
\tablewidth{18.5 cm}
\tabletypesize{\scriptsize}
\tablehead{
\colhead{Index}& \colhead{GLIMPSE ID}& \colhead{l ($\degr$)}& \colhead{b ($\degr$)}& \colhead{J}& \colhead{H}& \colhead{K}& \colhead{[3.6]}& \colhead{[4.5]}& \colhead{[5.8]}& \colhead{[8.0]}& \colhead{[24]}\\
\colhead{(1)}& \colhead{(2)}& \colhead{(3)}& \colhead{(4)}& \colhead{(5)}& \colhead{(6)}& \colhead{(7)}& \colhead{(8)}& \colhead{(9)}& \colhead{(10)}& \colhead{(11)}& \colhead{(12)}
}
\rotate
\startdata
  Y1&SSTGLMCG038.7727-00.5171&38.77265&-0.51719&16.839&14.860&13.814&13.206&12.880&12.606&11.979& 8.637\\
  Y2&SSTGLMCG038.7843-00.3593&38.78436&-0.35938&16.096&14.803&13.886&12.880&12.333&12.008&11.681& 8.108\\
  Y3&SSTGLMCG038.7863-00.5860&38.78636&-0.58613&14.749&13.770&12.777&11.527&11.009&10.556& 9.408& 6.023\\
  Y4&SSTGLMCG038.7901-00.3165&38.79005&-0.31655&16.580&15.083&14.111&13.218&12.461&12.639&11.943& 9.284\\
  Y5&SSTGLMCG038.8029-00.4189&38.80293&-0.41894&18.729&16.035&14.319&12.858&12.064&11.303&10.754& 6.635\\
  Y6&SSTGLMCG038.8092-00.2970&38.80917&-0.29705&16.902&15.393&14.342&13.004&12.150&11.539&10.748& 8.537\\
  Y7&SSTGLMCG038.8286-00.3209&38.82867&-0.32096&18.304&16.451&14.939&13.009&12.437&11.917&11.451& 8.230\\
  Y8&SSTGLMCG038.8324-00.4191&38.83249&-0.41917&\nodata&\nodata&\nodata&12.615&11.271&10.364& 9.708& 5.808\\
  Y9&SSTGLMCG038.8392-00.4590&38.83933&-0.45898&\nodata&\nodata&16.785&13.164&11.700& 9.697&\nodata& 5.257\\
 Y10&SSTGLMCG038.8394-00.5801&38.83943&-0.58015&\nodata&\nodata&\nodata&13.079&11.701&10.905&10.078& 6.136\\
 Y11&SSTGLMCG038.8399-00.5544&38.83993&-0.55451&\nodata&17.360&12.212& 9.336& 7.934& 7.102& 6.419& 2.440\\
 Y12&SSTGLMCG038.8399-00.3194&38.83996&-0.31949&15.416&13.812&12.086& 9.994& 9.137& 8.206& 6.639& 3.393\\
 Y13&SSTGLMCG038.8400-00.5465&38.83996&-0.54664&19.361&16.718&14.106&12.013&11.158&10.524&10.241& 7.017\\
 Y14&SSTGLMCG038.8412-00.5533&38.84124&-0.55336&\nodata&\nodata&\nodata&12.358&11.182&10.463& 9.997& 2.627\\
 Y15&SSTGLMCG038.8412-00.3865&38.84128&-0.38655&\nodata&\nodata&\nodata&12.193&11.455&10.745&10.184& 7.288\\
 Y16&SSTGLMCG038.8413-00.4279&38.84137&-0.42795&\nodata&\nodata&\nodata&13.068&12.215&11.429&10.888& 7.076\\
 Y17&SSTGLMCG038.8416-00.5732&38.84168&-0.57329&15.685&14.441&13.677&12.484&11.931&11.387&10.760& 8.382\\
 Y18&SSTGLMCG038.8425-00.4556&38.84258&-0.45563&\nodata&\nodata&\nodata&12.302&11.033&10.130& 8.967& 3.303\\
 Y19&SSTGLMCG038.8471-00.4295&38.84711&-0.42955&\nodata&\nodata&\nodata&13.180&10.878& 9.650& 8.659& 4.014\\
 Y20&SSTGLMCG038.8473-00.4281&38.84737&-0.42812&\nodata&\nodata&\nodata&13.512&12.010&11.510&10.764& 4.024\\
 Y21&SSTGLMCG038.8496-00.4263&38.84963&-0.42630&\nodata&\nodata&\nodata&13.095&12.043&11.484&10.786& 6.379\\
 Y22&SSTGLMCG038.8500-00.4611&38.85009&-0.46125&16.801&15.409&14.331&13.185&12.700&11.779&11.063& 8.430\\
 Y23&SSTGLMCG038.8510-00.4261&38.85104&-0.42625&\nodata&17.946&15.058&12.860&12.021&11.354&10.814& 7.433\\
 Y24&SSTGLMCG038.8515-00.5499&38.85151&-0.54995&\nodata&\nodata&\nodata&13.224&12.348&11.472&10.740& 6.690\\
 Y25&SSTGLMCG038.8528-00.5213&38.85283&-0.52130&\nodata&\nodata&\nodata&12.417&10.887&10.209& 9.602& 4.822\\
 Y26&SSTGLMCG038.8546-00.5827&38.85459&-0.58279&\nodata&17.122&15.876&13.598&12.862&12.125&11.629& 8.177\\
 Y27&SSTGLMCG038.8556-00.5113&38.85569&-0.51131&\nodata&\nodata&\nodata&14.030&12.404&11.242&10.138& 5.679\\
 Y28&SSTGLMCG038.8560-00.5322&38.85601&-0.53225&\nodata&\nodata&\nodata&14.022&12.703&11.609&11.053& 7.912\\
 Y29&SSTGLMCG038.8563-00.4272&38.85632&-0.42722&18.433&16.501&14.583&13.091&12.575&11.945&11.603& 7.343\\
 Y30&SSTGLMCG038.8569-00.4965&38.85691&-0.49665&\nodata&17.578&15.005&12.934&11.952&11.384&10.392& 6.663\\
 Y31&SSTGLMCG038.8586-00.5229&38.85864&-0.52295&17.488&14.963&13.485&11.313&10.568& 9.904&\nodata& 4.831\\
 Y32&SSTGLMCG038.8616-00.4239&38.86160&-0.42390&\nodata&\nodata&14.035&11.720&10.349&10.165& 9.357& 3.998\\
 Y33&SSTGLMCG038.8654-00.4686&38.86545&-0.46866&\nodata&\nodata&\nodata&14.632&13.412&\nodata&11.956& 7.823\\
 Y34&SSTGLMCG038.8661-00.4950&38.86615&-0.49513&15.394&14.136&13.340&12.196&11.553&11.228&10.548& 7.136\\
 Y35&SSTGLMCG038.8672-00.3364&38.86721&-0.33649&14.447&13.196&12.201&11.082&10.556&10.072& 8.952& 6.087\\
 Y36&SSTGLMCG038.8674-00.5602&38.86746&-0.56024&15.632&14.171&13.498&12.796&12.365&11.686&11.613& 8.463\\
 Y37&SSTGLMCG038.8712-00.4715&38.87121&-0.47161&\nodata&18.993&14.879&12.042&10.916& 9.865& 9.133& 6.269\\
 Y38&SSTGLMCG038.8729-00.3697&38.87297&-0.36980&15.961&14.633&13.768&12.514&12.025&11.359&10.896& 7.404\\
 Y39&SSTGLMCG038.8767-00.3622&38.87673&-0.36228&\nodata&\nodata&\nodata&12.696&11.761&11.217& 9.999& 6.715\\
 Y40&SSTGLMCG038.8787-00.3892&38.87875&-0.38924&\nodata&\nodata&\nodata&12.821&11.186&10.332& 9.439& 5.469\\
 Y41&SSTGLMCG038.8862-00.3861&38.88625&-0.38612&\nodata&\nodata&14.653&13.620&13.045&12.180&\nodata& 7.180\\
 Y42&SSTGLMCG038.8877-00.4184&38.88782&-0.41848&17.884&16.082&15.137&13.508&12.747&\nodata&11.390& 7.041\\
 Y43&SSTGLMCG038.8907-00.4560&38.89076&-0.45615&\nodata&\nodata&17.355&14.298&12.316&11.513&10.605& 5.219\\
 Y44&SSTGLMCG038.8917-00.4329&38.89171&-0.43306&16.856&15.485&14.659&13.625&12.964&12.116&\nodata& 7.290\\
 Y45&SSTGLMCG038.8923-00.3662&38.89229&-0.36628&19.180&16.564&15.097&13.623&12.791&12.087&10.929& 7.656\\
 Y46&SSTGLMCG038.8982-00.6693&38.89827&-0.66937&16.663&14.864&13.561&12.166&11.539&10.953&10.245& 7.603\\
 Y47&SSTGLMCG038.8999-00.4081&38.89989&-0.40819&17.422&15.596&14.210&13.039&12.165&11.946&\nodata& 6.876\\
 Y48&SSTGLMCG038.9060-00.4442&38.90600&-0.44436&14.607&13.120&12.368&11.726&11.188&10.356& 9.445& 3.468\\
 Y49&SSTGLMCG038.9061-00.4500&38.90613&-0.45013&18.743&16.789&14.445&12.389&11.535&11.023&10.354& 4.716\\
 Y50&SSTGLMCG038.9075-00.4693&38.90755&-0.46930&\nodata&\nodata&14.627&12.214&11.582&10.936&10.886& 6.582\\
 Y51&SSTGLMCG038.9108-00.3999&38.91086&-0.39991&\nodata&\nodata&\nodata&14.019&12.883&11.589&\nodata& 6.302\\
 Y52&SSTGLMCG038.9109-00.6495&38.91094&-0.64960&17.283&15.396&14.144&12.957&12.667&12.197&11.629& 8.250\\
 Y53&SSTGLMCG038.9126-00.5902&38.91265&-0.59021&\nodata&14.722&13.353&11.995&11.666&10.899&10.328& 8.081\\
 Y54&SSTGLMCG038.9127-00.3550&38.91269&-0.35511&\nodata&\nodata&15.136&12.770&11.729&11.103&\nodata& 5.492\\
 Y55&SSTGLMCG038.9138-00.4482&38.91380&-0.44828&17.465&15.595&13.831&12.525&11.926&11.288&10.759& 6.377\\
 Y56&SSTGLMCG038.9148-00.3587&38.91487&-0.35882&\nodata&\nodata&16.382&13.555&12.580&\nodata&\nodata& 5.321\\
 Y57&SSTGLMCG038.9174-00.4388&38.91742&-0.43893&16.144&14.482&13.084&11.493&10.844&10.479& 9.891& 6.011\\
 Y58&SSTGLMCG038.9183-00.4164&38.91839&-0.41652&19.126&16.938&15.700&13.736&12.888&12.026&\nodata& 6.132\\
 Y59&SSTGLMCG038.9194-00.4426&38.91947&-0.44273&16.942&15.400&14.453&13.206&12.534&11.890&10.998& 6.366\\
 Y60&SSTGLMCG038.9197-00.4468&38.91973&-0.44690&15.382&13.786&12.760&11.492&11.031&10.680& 9.968& 6.479\\
 Y61&SSTGLMCG038.9200-00.3846&38.92007&-0.38473&13.968&12.648&11.867&10.627&10.241& 9.756& 9.212& 5.590\\
 Y62&SSTGLMCG038.9202-00.3501&38.92021&-0.35012&\nodata&\nodata&\nodata&12.974&11.641&10.564&\nodata& 2.252\\
 Y63&SSTGLMCG038.9209-00.4163&38.92103&-0.41637&16.134&14.749&13.658&12.229&11.562&10.994&\nodata& 4.323\\
 Y64&SSTGLMCG038.9211-00.4172&38.92117&-0.41721&\nodata&\nodata&\nodata&13.371&11.859&10.341& 9.743& 4.255\\
 Y65&SSTGLMCG038.9214-00.3521&38.92140&-0.35216&\nodata&\nodata&\nodata&12.085& 9.684& 8.157&\nodata& 1.624\\
 Y66&SSTGLMCG038.9222-00.3566&38.92228&-0.35664&17.498&15.205&13.586&11.596&11.010&10.342&\nodata& 3.126\\
 Y67&SSTGLMCG038.9247-00.3685&38.92478&-0.36859&14.634&12.536&11.255& 9.865& 9.108& 8.603& 7.978& 5.614\\
 Y68&SSTGLMCG038.9248-00.3930&38.92489&-0.39305&16.144&14.618&13.612&12.512&11.747&11.044&\nodata& 5.612\\
 Y69&SSTGLMCG038.9258-00.4761&38.92581&-0.47617&\nodata&\nodata&16.333&13.508&12.620&11.873&11.229& 7.508\\
 Y70&SSTGLMCG038.9258-00.3694&38.92586&-0.36954&19.119&17.047&15.218&13.778&12.870&\nodata&\nodata& 5.249\\
 Y71&SSTGLMCG038.9271-00.4796&38.92717&-0.47969&\nodata&\nodata&\nodata&13.467&11.891&10.848&10.225& 6.067\\
 Y72&SSTGLMCG038.9274-00.4224&38.92754&-0.42245&18.731&16.743&15.276&13.849&12.836&\nodata&\nodata& 5.276\\
 Y73&SSTGLMCG038.9277-00.4432&38.92779&-0.44332&16.675&15.072&13.750&12.306&11.782&11.330&10.846& 6.820\\
 Y74&SSTGLMCG038.9295-00.3403&38.92956&-0.34034&18.071&16.178&14.659&13.017&12.038&11.584&10.850& 6.600\\
 Y75&SSTGLMCG038.9301-00.3063&38.93016&-0.30633&\nodata&\nodata&\nodata&13.362&11.290&10.324& 9.779& 5.832\\
 Y76&SSTGLMCG038.9308-00.3269&38.93081&-0.32695&\nodata&17.927&16.042&14.027&13.559&12.620&11.217& 8.454\\
 Y77&SSTGLMCG038.9311-00.3628&38.93120&-0.36287&\nodata&17.834&14.808&11.307& 9.966& 9.058& 8.259& 3.861\\
 Y78&SSTGLMCG038.9319-00.3439&38.93191&-0.34397&\nodata&\nodata&\nodata&13.815&12.722&11.711&\nodata& 4.286\\
 Y79&SSTGLMCG038.9319-00.3658&38.93199&-0.36589&\nodata&\nodata&17.522&13.768&12.894&11.870&\nodata& 4.936\\
 Y80&SSTGLMCG038.9325-00.2643&38.93260&-0.26443&\nodata&\nodata&16.375&12.772&11.432&10.364& 9.829& 6.653\\
 Y81&SSTGLMCG038.9330-00.3609&38.93295&-0.36102&17.877&14.159&11.461& 9.310& 8.167& 7.355& 6.600& 2.194\\
 Y82&SSTGLMCG038.9331-00.4660&38.93317&-0.46607&\nodata&17.428&13.666&11.207&10.225& 9.707& 9.298& 6.093\\
 Y83&SSTGLMCG038.9347-00.3349&38.93473&-0.33488&\nodata&\nodata&17.146&14.216&13.106&12.117&\nodata& 7.617\\
 Y84&SSTGLMCG038.9350-00.3598&38.93500&-0.35981&16.593&14.577&13.504&12.046&10.992&10.395& 9.849& 2.566\\
 Y85&SSTGLMCG038.9355-00.3638&38.93551&-0.36386&\nodata&\nodata&\nodata&12.483&11.019& 9.849& 9.165& 4.519\\
 Y86&SSTGLMCG038.9365-00.3144&38.93656&-0.31454&18.612&15.663&13.437&11.396&10.599& 9.980& 9.286& 6.925\\
 Y87&SSTGLMCG038.9371-00.3641&38.93712&-0.36427&15.592&14.084&12.941&11.782&11.267&10.751&10.100& 5.029\\
 Y88&SSTGLMCG038.9371-00.4497&38.93716&-0.44970&\nodata&\nodata&\nodata&13.624&12.838&12.010&10.804& 6.347\\
 Y89&SSTGLMCG038.9373-00.4527&38.93738&-0.45274&\nodata&\nodata&\nodata&13.128&12.120&11.599&10.786& 6.222\\
 Y90&SSTGLMCG038.9374-00.3596&38.93743&-0.35973&16.274&14.650&13.391&11.944&11.327&10.892& 9.811& 3.329\\
 Y91&SSTGLMCG038.9376-00.3625&38.93769&-0.36257&16.174&14.485&13.298&11.896&10.793&10.108& 9.323& 4.789\\
 Y92&SSTGLMCG038.9381-00.3504&38.93819&-0.35049&15.088&13.067&11.716&10.088& 9.522& 9.003&\nodata& 2.557\\
 Y93&SSTGLMCG038.9386-00.4509&38.93870&-0.45098&17.110&14.980&13.913&12.141&11.594&11.150&10.434& 6.391\\
 Y94&SSTGLMCG038.9388-00.4755&38.93885&-0.47552&\nodata&\nodata&\nodata&13.609&13.128&12.249&10.713& 6.807\\
 Y95&SSTGLMCG038.9402-00.3760&38.94021&-0.37608&\nodata&12.923&12.178&11.101&10.692&10.321& 9.837& 4.822\\
 Y96&SSTGLMCG038.9403-00.2857&38.94040&-0.28583&\nodata&18.125&15.581&13.168&12.309&11.715&11.354& 8.144\\
 Y97&SSTGLMCG038.9413-00.4804&38.94138&-0.48053&17.378&15.571&14.388&12.731&12.317&11.588&11.277& 7.032\\
 Y98&SSTGLMCG038.9418-00.6428&38.94180&-0.64287&\nodata&\nodata&\nodata&13.782&12.086&11.173&10.432& 5.637\\
 Y99&SSTGLMCG038.9446-00.3688&38.94468&-0.36886&16.969&15.232&14.281&11.778&10.853&10.244& 9.393& 5.841\\
Y100&SSTGLMCG038.9447-00.4359&38.94474&-0.43604&\nodata&\nodata&17.076&14.291&13.445&\nodata&\nodata& 5.856\\
Y101&SSTGLMCG038.9451-00.4409&38.94519&-0.44099&\nodata&\nodata&16.062&12.966&11.916&11.123&\nodata& 3.061\\
Y102&SSTGLMCG038.9455-00.4314&38.94550&-0.43147&16.024&14.677&13.655&12.413&11.738&11.363&10.946& 7.321\\
Y103&SSTGLMCG038.9456-00.4279&38.94563&-0.42797&\nodata&\nodata&15.791&12.536&11.186&10.313& 9.597& 6.217\\
Y104&SSTGLMCG038.9457-00.4609&38.94580&-0.46102&18.819&15.986&13.876&11.983&11.311&10.736&10.143& 4.772\\
Y105&SSTGLMCG038.9460-00.4357&38.94605&-0.43583&\nodata&17.305&14.601&12.364&11.333&10.528& 9.890& 6.593\\
Y106&SSTGLMCG038.9462-00.4624&38.94627&-0.46244&\nodata&\nodata&\nodata&15.008&12.845&12.000&11.338& 4.522\\
Y107&SSTGLMCG038.9465-00.4410&38.94660&-0.44109&\nodata&17.452&13.297&10.023& 8.759& 7.791& 6.864& 2.797\\
Y108&SSTGLMCG038.9480-00.4258&38.94805&-0.42593&17.119&15.538&14.323&13.020&12.361&11.824&11.007& 7.042\\
Y109&SSTGLMCG038.9481-00.4439&38.94816&-0.44397&\nodata&\nodata&\nodata&13.545&11.595&10.446& 9.766& 4.212\\
Y110&SSTGLMCG038.9485-00.4627&38.94847&-0.46285&\nodata&17.735&14.298&11.241& 9.861& 9.013& 8.255& 4.643\\
Y111&SSTGLMCG038.9486-00.4391&38.94868&-0.43911&\nodata&\nodata&\nodata&13.211&11.770&11.234&10.776& 3.874\\
Y112&SSTGLMCG038.9489-00.4283&38.94896&-0.42839&16.430&14.576&13.064&11.178&10.541&10.060& 9.665& 6.491\\
Y113&SSTGLMCG038.9494-00.4447&38.94943&-0.44475&14.368&13.735&13.508&13.054&12.138&10.412& 8.870& 4.676\\
Y114&SSTGLMCG038.9501-00.4649&38.95011&-0.46496&\nodata&\nodata&\nodata&14.469&12.521&11.676&10.698& 6.404\\
Y115&SSTGLMCG038.9513-00.4403&38.95141&-0.44038&18.054&16.103&14.968&13.272&12.609&11.954&11.305& 6.729\\
Y116&SSTGLMCG038.9518-00.4571&38.95185&-0.45719&16.763&15.409&14.479&13.481&12.831&12.345&11.543& 7.773\\
Y117&SSTGLMCG038.9528-00.4556&38.95288&-0.45572&17.142&15.785&14.550&13.034&12.310&11.982&11.982& 7.509\\
Y118&SSTGLMCG038.9551-00.4627&38.95520&-0.46281&17.899&15.342&13.618&11.508&10.855&10.287& 9.699& 3.910\\
Y119&SSTGLMCG038.9556-00.3524&38.95568&-0.35244&\nodata&\nodata&13.983&12.729&12.556&12.202&11.216& 8.180\\
Y120&SSTGLMCG038.9557-00.4653&38.95574&-0.46537&\nodata&\nodata&\nodata&14.199&12.307&11.515&10.703& 5.354\\
Y121&SSTGLMCG038.9565-00.4465&38.95656&-0.44658&\nodata&\nodata&\nodata&12.950&12.248&11.577&10.951& 8.150\\
Y122&SSTGLMCG038.9570-00.4625&38.95698&-0.46260&\nodata&15.127&12.895&10.755& 9.632& 8.789& 7.734& 3.267\\
Y123&SSTGLMCG038.9570-00.4695&38.95708&-0.46951&\nodata&\nodata&\nodata&14.738&12.914&12.030&11.407& 7.195\\
Y124&SSTGLMCG038.9594-00.4206&38.95943&-0.42071&\nodata&17.475&13.759&10.440& 9.242& 8.309& 7.570& 5.472\\
Y125&SSTGLMCG038.9623-00.4576&38.96236&-0.45767&\nodata&15.634&14.384&12.771&12.272&11.800&11.246& 6.927\\
Y126&SSTGLMCG038.9624-00.4613&38.96245&-0.46140&\nodata&\nodata&17.212&14.310&13.467&\nodata&\nodata& 6.477\\
Y127&SSTGLMCG038.9625-00.4741&38.96251&-0.47419&\nodata&\nodata&17.814&13.677&11.983&11.162&10.589& 6.487\\
Y128&SSTGLMCG038.9636-00.4760&38.96367&-0.47615&\nodata&\nodata&16.478&13.124&11.575&10.792&10.258& 6.814\\
Y129&SSTGLMCG038.9650-00.3061&38.96504&-0.30617&\nodata&18.256&15.694&13.585&12.837&12.060&11.282& 7.581\\
Y130&SSTGLMCG038.9652-00.4741&38.96527&-0.47423&\nodata&\nodata&14.008&11.000&10.318& 9.509& 9.495& 6.818\\
Y131&SSTGLMCG038.9665-00.4139&38.96653&-0.41399&\nodata&\nodata&\nodata&15.537&13.593&11.660&10.877& 7.219\\
Y132&SSTGLMCG038.9667-00.2415&38.96670&-0.24154&\nodata&\nodata&\nodata&13.557&12.212&11.358&10.642& 6.712\\
Y133&SSTGLMCG038.9671-00.4629&38.96714&-0.46299&\nodata&\nodata&18.103&13.372&11.065& 9.685& 8.503& 3.421\\
Y134&SSTGLMCG038.9679-00.4670&38.96792&-0.46701&14.844&14.096&13.776&13.480&12.832&\nodata&\nodata& 7.601\\
Y135&SSTGLMCG038.9681-00.2713&38.96818&-0.27141&\nodata&16.689&14.983&13.313&12.412&11.055&\nodata& 5.613\\
Y136&SSTGLMCG038.9696-00.4065&38.96961&-0.40651&\nodata&\nodata&\nodata&14.798&13.659&\nodata&11.727& 7.580\\
Y137&SSTGLMCG038.9712-00.3579&38.97124&-0.35796&15.940&14.284&13.024&11.288&10.310& 9.710& 8.523& 4.973\\
Y138&SSTGLMCG038.9732-00.5394&38.97329&-0.53944&\nodata&\nodata&\nodata&12.384&11.483&10.844&10.489& 6.945\\
Y139&SSTGLMCG038.9737-00.2857&38.97380&-0.28576&\nodata&\nodata&17.005&13.904&12.303&11.429&10.955& 7.375\\
Y140&SSTGLMCG038.9756-00.2829&38.97568&-0.28294&\nodata&\nodata&\nodata&14.064&12.818&11.852&11.251& 8.073\\
Y141&SSTGLMCG038.9771-00.4591&38.97717&-0.45917&\nodata&\nodata&\nodata&14.824&13.135&12.066&11.211& 6.578\\
Y142&SSTGLMCG038.9791-00.2670&38.97923&-0.26711&18.103&15.520&13.704&11.430&10.620& 9.892& 8.987& 5.603\\
Y143&SSTGLMCG038.9793-00.4826&38.97938&-0.48268&\nodata&\nodata&15.642&12.795&12.200&11.520&\nodata& 8.441\\
Y144&SSTGLMCG038.9814-00.2720&38.98147&-0.27209&17.381&15.121&13.879&12.583&11.214&\nodata&\nodata& 4.467\\
Y145&SSTGLMCG038.9858-00.5569&38.98586&-0.55702&\nodata&17.764&15.247&12.792&11.925&11.195&10.145& 5.764\\
Y146&SSTGLMCG038.9867-00.3409&38.98671&-0.34100&16.591&15.137&14.601&13.086&12.705&12.341&11.802& 8.866\\
Y147&SSTGLMCG038.9872-00.4934&38.98723&-0.49353&\nodata&\nodata&16.930&14.732&13.292&\nodata&\nodata& 7.770\\
Y148&SSTGLMCG038.9882-00.2654&38.98818&-0.26555&17.005&15.442&14.301&13.025&12.455&12.088&11.113& 6.870\\
Y149&SSTGLMCG038.9915-00.2670&38.99152&-0.26712&17.424&15.735&14.710&13.941&13.334&\nodata&12.122& 7.607\\
Y150&SSTGLMCG038.9982-00.5070&38.99830&-0.50702&17.332&15.088&12.495& 9.760& 8.541& 7.650& 6.822& 3.454\\
Y151&SSTGLMCG039.0022-00.5148&39.00229&-0.51484&15.918&14.341&13.584&12.564&12.232&11.690&11.206& 7.117\\
Y152&SSTGLMCG039.0107-00.4447&39.01079&-0.44479&14.630&13.722&12.850&11.629&11.168&10.552& 9.819& 8.162\\
Y153&SSTGLMCG039.0112-00.4421&39.01127&-0.44214&14.836&13.618&13.010&12.693&12.549&12.236&11.566& 8.533\\
Y154&SSTGLMCG039.0186-00.3259&39.01862&-0.32593&13.435&12.876&12.489&12.370&12.091&11.725&11.305& 8.643\\
Y155&SSTGLMCG039.0362-00.6522&39.03620&-0.65228&17.078&15.549&14.406&13.257&12.554&12.083&11.643& 7.421\\
Y156&SSTGLMCG039.0420-00.4910&39.04206&-0.49110&12.945&12.219&11.877&11.592&11.403&11.294&10.774& 5.754\\
Y157&SSTGLMCG039.0423-00.2613&39.04240&-0.26134&13.741&13.185&12.750&12.530&12.437&12.034&11.606& 7.678\\
Y158&SSTGLMCG039.0446-00.4460&39.04465&-0.44612&15.322&13.738&12.913&12.331&12.257&12.051&11.326& 9.151\\
Y159&SSTGLMCG039.0463-00.5163&39.04630&-0.51636&\nodata&\nodata&\nodata&13.525&10.685& 9.814& 8.711& 4.450\\
Y160&SSTGLMCG039.0496-00.5965&39.04968&-0.59658&15.031&12.689&11.727&10.286& 9.580& 8.729& 7.733& 5.427\\
Y161&SSTGLMCG039.0528-00.4631&39.05282&-0.46315&14.221&13.413&13.028&12.773&12.693&12.497&11.657& 9.153\\
Y162&SSTGLMCG039.0572-00.4051&39.05726&-0.40519&\nodata&\nodata&15.965&13.621&12.831&11.198& 9.827& 6.595\\
\enddata
\end{deluxetable}

\clearpage
\begin{deluxetable}{lrcccccccccccc}
\rotate
\tablecaption{Candidate YSO Parameters \label{ysoparams}}
\tablewidth{18.5 cm}
\tabletypesize{\scriptsize}
\tablehead{
\colhead{Index}& \colhead{Bands}& \colhead{\# Fits}& \colhead{$\chi^{2}$/n$_{data}$}& \multicolumn{2}{c}{M$_{*}$(\msun)}& \multicolumn{2}{c}{log$_{10}$(M$_{disk}$)/\msun}& \multicolumn{2}{c}{log$_{10}$($\dot{M}$/\msun~yr$^{-1}$)}& \multicolumn{2}{c}{Integer Stage}& \colhead{F$_{stage}$}& \colhead{$\sigma_{stage}$}\\
\colhead{}& \colhead{}& \colhead{}& \colhead{BEST}& \colhead{BEST}& \colhead{AVG}& \colhead{BEST}& \colhead{AVG}& \colhead{BEST}& \colhead{AVG}& \colhead{BEST}& \colhead{AVG}& \colhead{AVG}& \colhead{AVG}\\
\colhead{(1)}& \colhead{(2)}& \colhead{(3)}& \colhead{(4)}& \colhead{(5)}& \colhead{(6)}& \colhead{(7)}& \colhead{(8)}& \colhead{(9)}& \colhead{(10)}& \colhead{(11)}& \colhead{(12)}& \colhead{(13)}& \colhead{(14)}
}
\startdata
  Y1&11111113& 2634&  0.14& 0.7& 1.6& -3.35& -2.80& -6.47& -6.36& II&amb&2.02&0.39\\
  Y2&11111113& 1232&  0.59& 1.7& 1.8& -1.57& -1.98& -6.03& -6.87& II& II&2.01&0.06\\
  Y3&11111111&  101&  0.16& 2.6& 2.5& -2.60& -2.79&  0.00& -8.43& II& II&2.00&0.00\\
  Y4&11111111& 1540&  1.12& 1.9& 1.8& -3.88& -2.09&  0.00& -7.12& II& II&2.02&0.07\\
  Y5&11111111&  261&  0.36& 0.6& 1.3& -2.77& -1.96& -5.48& -5.29&  I&  I&1.27&0.39\\
  Y6&11111111&  109&  0.85& 2.5& 2.3& -1.51& -1.61&  0.00& -7.34& II& II&2.00&0.00\\
  Y7&11111111&   21&  0.32& 2.2& 2.7& -1.93& -2.16& -4.73& -4.98&  I&amb&1.44&0.49\\
  Y8&00011111&   52&  0.08& 4.1& 3.2& -0.59& -0.92& -3.86& -3.80&  I&  I&1.05&0.10\\
  Y9&00111101&   28&  3.42& 1.2& 1.8& -1.40& -1.58& -4.16& -3.86&  I&  I&1.13&0.23\\
 Y10&00011111&  106&  0.19& 4.1& 3.1& -0.59& -1.08& -3.86& -4.07&  I&  I&1.18&0.29\\
 Y11&01111111&    8&  7.61& 9.2& 8.8& -2.14& -1.82&  0.00& -7.32& II& II&2.00&0.00\\
 Y12&11111111&   40&  1.60& 4.1& 4.1& -1.95& -1.99&  0.00& -4.72& II& II&1.80&0.32\\
 Y13&11111111&  171&  1.54& 2.3& 2.6& -1.30& -1.43& -4.00& -4.58&  I&amb&1.62&0.47\\
 Y14&00011113&  947&  0.00& 6.4& 3.9& -2.32& -1.41& -3.75& -3.99&  I&  I&1.02&0.03\\
 Y15&00011111&  882&  0.10& 1.5& 3.3& -1.94& -1.75& -4.76& -5.86&  I& II&1.95&0.09\\
 Y16&00011113& 1320&  0.11& 0.8& 2.9& -2.19& -1.86& -5.41& -5.29&  I& II&1.87&0.23\\
 Y17&11111113&  135&  0.16& 2.3& 2.2& -1.68& -1.95&  0.00& -8.04& II& II&2.00&0.01\\
 Y18&00011111&  478&  0.03& 5.7& 3.2& -2.72& -1.35& -5.03& -4.35&  I&  I&1.06&0.11\\
 Y19&00011111&   65&  0.87& 1.3& 0.9& -1.57& -1.73& -5.17& -5.08&  I&  I&1.00&0.00\\
 Y20&00011113&  858&  1.11& 5.2& 3.0& -2.32& -1.52& -4.54& -4.16&  I&  I&1.08&0.15\\
 Y21&00011111&  568&  0.24& 1.6& 2.4& -1.12& -1.52& -4.77& -4.40&  I&  I&1.17&0.29\\
 Y22&11111111&  129&  0.31& 1.5& 1.6& -1.92& -1.82&  0.00& -8.19& II& II&2.00&0.00\\
 Y23&01111113&  670&  0.17& 2.9& 3.2& -2.64& -2.06&  0.00& -5.81& II& II&1.97&0.07\\
 Y24&00011111&  498&  0.11& 0.6& 2.1& -2.77& -1.63& -5.48& -4.82&  I&  I&1.29&0.41\\
 Y25&00011111&   89&  0.80& 6.0& 3.8& -1.58& -1.11& -3.01& -3.54&  I&  I&1.00&0.01\\
 Y26&01111111&   11&  0.49& 0.8& 0.9& -1.83& -1.85& -4.88& -4.86&  I&  I&1.05&0.10\\
 Y27&00011111&  242&  0.18& 0.5& 0.8& -3.32& -1.94& -5.50& -4.75&  I&  I&1.07&0.13\\
 Y28&00011113&   10&  0.24& 0.6& 1.6& -3.24& -2.65& -3.66& -3.84&  I&amb&1.34&0.45\\
 Y29&11111113&  436&  0.59& 0.6& 2.0& -2.15& -2.23& -4.82& -5.62&  I& II&1.84&0.29\\
 Y30&01111111&  284&  0.12& 1.7& 2.2& -2.52& -1.70& -5.72& -5.15&  I&amb&1.52&0.50\\
 Y31&11111101&  129&  0.97& 1.8& 2.6& -2.19& -1.51& -5.00& -4.28&  I&  I&1.15&0.25\\
 Y32&00111111&  497&  2.05& 5.8& 4.7& -1.54& -1.48& -5.69& -4.21& II&  I&1.14&0.24\\
 Y33&00011011&  993&  0.06& 1.8& 1.4& -3.69& -1.88& -3.63& -4.77&  I&  I&1.32&0.44\\
 Y34&11111111&  311&  0.60& 2.1& 2.0& -1.07& -1.59& -8.25& -6.25& II& II&1.92&0.15\\
 Y35&11111111&  278&  0.28& 2.8& 2.8& -4.23& -2.75&  0.00& -7.51& II& II&2.00&0.01\\
 Y36&11111111& 1743&  0.85& 1.6& 1.5& -2.94& -2.39& -6.88& -6.08& II& II&1.78&0.34\\
 Y37&01111111&  156&  0.68& 4.4& 4.1& -3.21& -1.76&  0.00&  0.00& II& II&2.00&0.00\\
 Y38&11111111&  132&  0.74& 2.5& 2.1& -1.06& -1.38& -6.29& -6.52& II& II&1.99&0.01\\
 Y39&00011111&  913&  0.10& 3.5& 3.0& -3.99& -2.12&  0.00& -5.82& II& II&1.94&0.17\\
 Y40&00011111&   60&  0.44& 4.9& 3.1& -1.08& -1.19&  0.00& -4.22& II&amb&1.42&0.49\\
 Y41&00111103& 1950&  0.01& 3.4& 1.5& -2.58& -1.99& -6.68& -5.74& II&amb&1.79&0.43\\
 Y42&11111013&  168&  0.66& 0.3& 1.2& -2.20& -2.29& -6.73& -4.96& II&amb&1.66&0.54\\
 Y43&00111111&  233&  1.12& 1.9& 1.6& -1.51& -1.78& -5.09& -4.33&  I&  I&1.02&0.04\\
 Y44&11111103&  264&  0.02& 1.0& 1.4& -1.53& -1.96&  0.00& -6.61& II& II&1.99&0.22\\
 Y45&11111111&  216&  0.47& 2.0& 1.6& -1.53& -1.68&  0.00& -5.56& II& II&1.80&0.33\\
 Y46&11111111&  323&  0.23& 2.3& 2.5& -2.01& -2.15&  0.00& -7.52& II& II&2.00&0.00\\
 Y47&11111103& 1042&  0.67& 1.9& 2.3& -4.36& -2.17&  0.00& -6.45& II& II&2.14&0.31\\
 Y48&11111113&  518&  0.31& 1.4& 1.9& -3.68& -1.80& -4.52& -4.75&  I&  I&1.19&0.31\\
 Y49&11111113&  165&  0.86& 3.2& 1.9& -2.91& -1.68& -4.34& -4.36&  I&  I&1.04&0.08\\
 Y50&00111111&   79&  0.69& 4.0& 2.9& -1.40& -1.43& -4.42& -4.32&  I&  I&1.09&0.17\\
 Y51&00011103&  663&  0.01& 0.8& 2.4& -1.65& -1.92& -4.82& -4.58&  I&amb&1.69&0.56\\
 Y52&11111113& 2756&  0.12& 0.8& 2.0& -3.98& -2.18& -5.86& -6.39&  I& II&2.00&0.16\\
 Y53&01111111&  411&  0.60& 2.3& 2.8& -2.01& -2.26&  0.00& -7.11& II& II&2.08&0.15\\
 Y54&00111103& 2269&  0.05& 4.0& 3.3& -3.90& -1.94& -7.49& -5.22& II& II&1.96&0.24\\
 Y55&11111113&  715&  0.42& 2.2& 2.3& -1.48& -2.07&  0.00& -6.31& II& II&1.96&0.09\\
 Y56&00111003& 2268&  0.00& 3.3& 2.2& -7.07& -1.81&  0.00& -4.71&III&amb&1.45&0.54\\
 Y57&11111113&  297&  0.39& 2.9& 2.8& -2.64& -2.37&  0.00& -6.49& II& II&1.96&0.07\\
 Y58&11111103&  381&  0.26& 1.5& 1.2& -1.94& -2.17& -4.76& -4.75&  I&  I&1.28&0.42\\
 Y59&11111113&  267&  0.12& 0.3& 1.4& -1.71& -1.89& -5.58& -4.93&  I& II&1.71&0.43\\
 Y60&11111111&  497&  0.31& 3.9& 2.7& -1.49& -1.52& -6.21& -5.53& II& II&1.79&0.33\\
 Y61&11111113&  450&  0.10& 4.1& 3.3& -0.91& -1.93& -7.55& -6.07& II& II&1.99&0.03\\
 Y62&00011103& 6803&  0.00& 3.5& 3.2& -1.63& -1.49& -4.48& -4.14&  I&  I&1.36&0.50\\
 Y63&11111103&  604&  0.05& 2.2& 3.2& -5.54& -2.30&  0.00& -5.69& II&amb&2.42&0.55\\
 Y64&00011111&  395&  0.07& 3.8& 2.4& -2.50& -1.57& -4.21& -3.85&  I&  I&1.00&0.00\\
 Y65&00011101&  276&  0.12& 8.5& 5.8& -1.42& -1.00& -2.73& -3.60&  I&  I&1.08&0.14\\
 Y66&11111103& 1244&  0.17& 1.0& 2.5& -1.66& -1.80& -5.13& -4.83&  I&amb&1.47&0.57\\
 Y67&11111111&  328&  0.44& 4.0& 3.8& -3.45& -2.03&  0.00& -9.80& II& II&2.00&0.00\\
 Y68&11111103&  675&  0.06& 2.1& 2.4& -1.81& -1.71&  0.00& -6.30& II& II&2.04&0.22\\
 Y69&00111113&  346&  0.03& 1.5& 2.8& -3.07& -2.19& -3.86& -5.05&  I& II&1.88&0.22\\
 Y70&11111003& 1452&  0.15& 4.6& 1.4& -1.08& -2.24&  0.00& -5.23& II&amb&1.59&0.66\\
 Y71&00011111&   58&  0.07& 4.1& 3.0& -0.59& -0.99& -3.86& -3.74&  I&  I&1.01&0.02\\
 Y72&11111003&  822&  0.03& 2.7& 1.4& -3.96& -2.25&  0.00& -4.82& II&amb&1.47&0.58\\
 Y73&11111113&  456&  0.16& 3.8& 2.3& -4.39& -2.04&  0.00& -6.94& II& II&1.98&0.04\\
 Y74&11111111&   70&  0.63& 0.6& 1.2& -2.77& -1.98& -5.48& -5.27&  I&  I&1.19&0.31\\
 Y75&00011111&    8&  1.55& 1.5& 3.1& -2.00& -1.34& -3.89& -3.33&  I&  I&1.00&0.00\\
 Y76&01111113&  195&  0.83& 2.7& 2.8& -6.45& -3.00&  0.00& -7.88&III&III&2.75&0.38\\
 Y77&01111113&   24&  0.50& 3.9& 3.7& -2.22& -1.56&  0.00& -4.92& II& II&1.84&0.26\\
 Y78&00011103& 5475&  0.00& 0.1& 2.0& -3.91& -1.74& -5.97& -4.39&  I&  I&1.26&0.41\\
 Y79&00111103&  714&  0.23& 5.5& 2.4& -3.51& -1.69& -3.91& -4.23&  I&  I&1.14&0.24\\
 Y80&00111111&   13&  1.25& 5.5& 5.5& -3.75& -3.04&  0.00&  0.00& II& II&2.00&0.00\\
 Y81&11111111&  250&  1.01& 6.3& 6.0& -1.97& -1.47& -6.95& -4.05& II& II&1.88&0.21\\
 Y82&01111113&  102&  0.82& 4.3& 4.4& -1.23& -1.43&  0.00&  0.00& II& II&2.00&0.00\\
 Y83&00111101&  302&  0.05& 1.3& 2.2& -3.40& -1.86& -4.98& -4.79&  I&amb&1.66&0.49\\
 Y84&11111113&  227&  0.20& 5.3& 3.9& -1.13& -1.36& -4.38& -4.38&  I&  I&1.09&0.17\\
 Y85&00011113&   86&  0.00& 2.4& 2.1& -1.71& -1.42& -4.02& -3.87&  I&  I&1.07&0.13\\
 Y86&11111111&  428&  0.13& 5.1& 3.4& -3.17& -2.51&  0.00& -8.65& II& II&2.00&0.00\\
 Y87&11111113& 1067&  0.06& 2.3& 2.6& -2.01& -1.82&  0.00& -5.72& II& II&1.88&0.23\\
 Y88&00011113& 2189&  0.00& 0.4& 2.2& -2.34& -1.95& -6.23& -5.38&  I&amb&1.80&0.53\\
 Y89&00011113& 3450&  0.05& 3.1& 2.9& -2.12& -1.91&  0.00& -5.25& II& II&1.87&0.28\\
 Y90&11111113& 1058&  0.08& 4.3& 2.9& -4.92& -2.36& -6.77& -5.54& II& II&2.10&0.33\\
 Y91&11111111&   21&  0.55& 0.3& 1.2& -1.66& -1.64& -4.32& -4.21&  I&  I&1.00&0.00\\
 Y92&11111103& 1571&  0.02& 3.5& 3.8& -2.90& -1.98&  0.00& -5.35& II& II&1.97&0.18\\
 Y93&11111113&  475&  0.91& 2.3& 2.4& -1.56& -1.80& -5.06& -5.74&  I& II&1.74&0.39\\
 Y94&00011113& 1009&  0.00& 0.6& 2.1& -2.64& -2.02& -5.62& -5.80&  I&amb&1.95&0.39\\
 Y95&01111113&  970&  0.05& 2.9& 3.0& -2.37& -1.65&  0.00& -5.11& II&amb&1.61&0.48\\
 Y96&01111111&   10&  0.17& 1.0& 1.1& -1.57& -1.58& -4.38& -4.39&  I&  I&1.02&0.04\\
 Y97&11111113&  185&  0.78& 3.8& 2.1& -3.24& -1.93& -5.79& -5.52& II&amb&1.56&0.49\\
 Y98&00011111&  116&  0.49& 1.7& 2.2& -1.98& -1.41& -4.31& -4.00&  I&  I&1.01&0.01\\
 Y99&11111113&   75&  3.23& 2.4& 2.2& -2.00& -2.33&  0.00& -5.03& II& II&1.93&0.14\\
Y100&00111003& 2562&  0.01& 1.2& 1.4& -2.96& -1.96& -4.79& -4.83&  I&  I&1.24&0.39\\
Y101&00111103& 1578&  0.00& 1.8& 3.1& -1.39& -1.63& -4.28& -4.40&  I&amb&1.42&0.56\\
Y102&11111113&  242&  0.50& 2.2& 2.2& -1.48& -2.12&  0.00& -7.48& II& II&1.99&0.02\\
Y103&00111111&   57&  0.62& 4.6& 4.1& -1.10& -1.36&  0.00& -5.20& II& II&1.97&0.09\\
Y104&11111113&  805&  0.03& 1.3& 2.5& -1.24& -1.70& -4.80& -4.90&  I&amb&1.54&0.50\\
Y105&01111113&  358&  0.38& 4.5& 3.2& -3.17& -2.10&  0.00& -6.88& II& II&2.00&0.01\\
Y106&00011113&  325&  1.48& 5.5& 3.1& -2.66& -1.28& -3.20& -3.69&  I&  I&1.00&0.00\\
Y107&01111111&   65&  0.32& 5.8& 6.2& -2.20& -1.55& -8.14& -4.71& II& II&1.95&0.10\\
Y108&11111111&   74&  0.25& 2.2& 1.9& -2.17& -1.82&  0.00& -5.18& II& II&1.77&0.35\\
Y109&00011111&  118&  0.41& 6.8& 4.7& -2.50& -1.35& -3.34& -3.50&  I&  I&1.00&0.00\\
Y110&01111111&   70&  0.89& 5.1& 4.7& -0.82& -1.12& -3.03& -3.54&  I&amb&1.61&0.48\\
Y111&00011111&  203&  0.87& 5.3& 4.1& -2.13& -1.68& -4.31& -4.36&  I&  I&1.00&0.00\\
Y112&11111111&  211&  1.41& 3.0& 2.9& -1.75& -1.62&  0.00& -5.95& II& II&1.94&0.12\\
Y113&11111113&   32&  2.61& 3.0& 3.0& -4.83& -4.78&  0.00&  0.00& II& II&2.00&0.00\\
Y114&00011111&   91&  1.77& 2.4& 1.8& -1.71& -1.37& -4.02& -4.31&  I&  I&1.09&0.17\\
Y115&11111113&  267&  0.36& 2.5& 1.6& -1.06& -1.68& -6.29& -4.85& II&amb&1.54&0.50\\
Y116&11111113&  534&  0.11& 1.5& 1.4& -1.92& -2.08&  0.00& -6.41& II& II&1.92&0.18\\
Y117&11111113&   19&  2.01& 3.8& 3.2& -3.24& -2.13& -5.79& -5.78& II& II&1.79&0.33\\
Y118&11111113&  528&  0.24& 1.5& 2.4& -1.64& -1.60& -4.99& -4.52&  I&  I&1.24&0.36\\
Y119&00111113& 2585&  0.20& 2.9& 2.4& -6.16& -2.62&  0.00& -6.55&III&amb&2.37&0.53\\
Y120&00011113&  142&  1.65& 4.1& 2.3& -0.59& -1.27& -3.86& -3.87&  I&  I&1.03&0.06\\
Y121&00011113& 1091&  0.12& 3.3& 2.8& -2.15& -2.11& -6.38& -6.75& II& II&2.01&0.04\\
Y122&01111111&   60&  0.48& 0.4& 2.6& -1.41& -1.69& -5.53& -4.52&  I&  I&1.02&0.03\\
Y123&00011113&   20&  1.27& 0.6& 1.7& -3.24& -1.58& -3.66& -3.84&  I&  I&1.02&0.03\\
Y124&01111111&   48&  1.37& 5.1& 5.7& -3.33& -3.12&  0.00&  0.00& II& II&2.00&0.00\\
Y125&01111113& 2297&  0.06& 0.6& 2.2& -1.98& -1.97& -5.13& -5.54&  I& II&1.78&0.37\\
Y126&00111003& 1207&  0.03& 1.7& 1.6& -3.10& -1.95& -4.30& -4.79&  I&  I&1.30&0.45\\
Y127&00111111&   15&  1.08& 2.6& 3.4& -1.15& -0.92& -3.96& -3.69&  I&  I&1.00&0.00\\
Y128&00111111&   34&  2.38& 2.8& 3.3& -1.37& -1.42&  0.00& -4.37& II& II&1.72&0.41\\
Y129&01111113&  912&  0.14& 0.8& 2.5& -2.19& -1.90& -5.41& -6.18&  I& II&1.93&0.14\\
Y130&00111113&   27&  5.21& 4.3& 4.3& -1.23& -1.40&  0.00&  0.00& II& II&2.00&0.00\\
Y131&00011111&   14&  1.21& 1.8& 1.5& -1.58& -1.82& -3.26& -3.40&  I&  I&1.00&0.01\\
Y132&00011111&  104&  0.26& 1.8& 2.3& -2.46& -1.62& -4.22& -4.26&  I&  I&1.13&0.23\\
Y133&00111111&   93&  0.44& 3.1& 2.0& -2.38& -1.35& -4.54& -4.70&  I&  I&1.00&0.00\\
Y134&11111003&  172&  0.48& 1.9& 2.4& -5.15& -5.14&  0.00&  0.00& II&III&2.70&0.42\\
Y135&01111101&  518&  0.22& 0.2& 1.0& -2.60& -1.95& -5.48& -4.89&  I&  I&1.15&0.25\\
Y136&00011011&  511&  0.00& 1.3& 1.3& -2.67& -2.01& -5.07& -5.03&  I&  I&1.29&0.42\\
Y137&11111111&   39&  1.20& 3.2& 3.1& -1.23& -1.49& -4.88& -5.07&  I&amb&1.49&0.50\\
Y138&00011111&  119&  0.26& 1.5& 3.0& -3.07& -1.45& -3.86& -4.27&  I&  I&1.27&0.39\\
Y139&00111111&   31&  2.04& 1.0& 1.9& -1.18& -1.49& -4.62& -4.18&  I&  I&1.24&0.36\\
Y140&00011113&   55&  0.38& 4.5& 3.0& -3.17& -2.19&  0.00& -4.95& II& II&1.83&0.30\\
Y141&00011111&  401&  0.20& 1.7& 1.1& -1.98& -1.66& -4.31& -4.39&  I&  I&1.06&0.12\\
Y142&11111111&   95&  0.67& 1.8& 3.2& -2.46& -2.47& -4.22& -4.71&  I&amb&1.48&0.50\\
Y143&00111103&  219&  0.73& 4.4& 3.7& -6.35& -2.69&  0.00&  0.00&III&amb&2.54&0.50\\
Y144&11111003&  421&  0.76& 3.2& 2.6& -1.09& -1.59& -7.74& -5.27& II&amb&1.77&0.54\\
Y145&01111111&  125&  0.12& 3.3& 2.2& -3.02& -1.72&  0.00& -4.65& II&amb&1.53&0.50\\
Y146&11111113&  168&  1.53& 1.2& 1.9& -2.21& -2.13& -8.16& -8.03& II& II&2.03&0.06\\
Y147&00111003&  783&  0.19& 5.5& 2.4& -7.36& -1.93&  0.00& -5.36&III&amb&1.99&0.44\\
Y148&11111113&  655&  0.09& 2.0& 2.2& -4.51& -2.40&  0.00& -6.36& II& II&2.16&0.30\\
Y149&11111013& 5656&  0.06& 0.3& 1.1& -2.50& -2.35& -5.14& -5.91&  I& II&1.79&0.41\\
Y150&11111111&   48&  0.96& 4.5& 4.7& -1.92& -1.73&  0.00& -5.29& II& II&1.99&0.02\\
Y151&11111111&  714&  0.22& 2.3& 1.5& -4.68& -2.16& -5.38& -5.47&  I&amb&1.52&0.50\\
Y152&11111111&  173&  0.94& 2.2& 2.6& -4.62& -3.26&  0.00&  0.00& II& II&2.25&0.37\\
Y153&11111113& 1161&  0.04& 2.3& 2.3& -5.26& -3.56& -6.48& -6.99& II&amb&2.46&0.51\\
Y154&11111113&   64&  0.47& 3.1& 3.1& -7.66& -7.61&  0.00&  0.00&III&III&3.00&0.00\\
Y155&11111113&  498&  0.36& 2.1& 1.6& -1.24& -2.01& -6.23& -6.22& II& II&1.85&0.29\\
Y156&11111111&   19&  0.93& 3.6& 3.6& -4.85& -4.84&  0.00& -8.86& II& II&2.00&0.00\\
Y157&11111111&   42&  0.96& 3.9& 3.6& -7.17& -6.42&  0.00& -8.72&III&III&2.99&0.01\\
Y158&11111113&  353&  0.21& 3.2& 3.0& -6.03& -3.74& -8.29& -7.47&III&III&2.72&0.41\\
Y159&00011111&   11&  4.76& 0.5& 0.6& -1.96& -1.84& -5.54& -5.44&  I&  I&1.00&0.00\\
Y160&11111111&  183&  2.71& 3.2& 3.8& -1.56& -1.98&  0.00& -8.01& II& II&2.06&0.12\\
Y161&11111113&  161&  0.29& 3.1& 2.8& -7.66& -6.61&  0.00& -8.01&III&III&2.97&0.06\\
Y162&00111111&   27&  2.03& 3.9& 3.8& -1.60& -1.59& -7.56& -5.16& II& II&1.97&0.05\\
\enddata
\end{deluxetable}

\clearpage

\clearpage

\clearpage

\clearpage

\clearpage

\clearpage

\clearpage

\clearpage

\clearpage

\clearpage

\clearpage

\clearpage

\clearpage
  
\clearpage

\clearpage

\clearpage

\clearpage

\clearpage


\begin{thebibliography}{}
\bibitem[Andre et al.(2000)]{ad00} Andre, P., Ward-Thompson, D., \& Barsony, M.\ 2000, Protostars and Planets IV, 59 
\bibitem[Allen et al.(2004)]{al04} Allen, L.~E., Calvet, N., D'Alessio, P., et al.\ 2004, \apjs, 154, 363 
\bibitem[Alexander \& Kobulnicky(2012)]{al12} Alexander, M.~J., \& Kobulnicky, H.~A.\ 2012, \apjl, 755, L30 
\bibitem[Arvidsson et al.(2010)]{ar10} Arvidsson, K., Kerton, C.~R., Alexander, M.~J., Kobulnicky, H.~A., \& Uzpen, B.\ 2010, \aj, 140, 462 
\bibitem[Ballesteros-Paredes et al.(2011)]{ba11} Ballesteros-Paredes, J., Hartmann, L.~W., V{\'a}zquez-Semadeni, E., Heitsch, F., \& Zamora-Avil{\'e}s, M.~A.\ 2011, \mnras, 411, 65 
\bibitem[Balser et al.(2011)]{balser2011} Balser, D.~S., Rood, R.~T., Bania, T.~M., \& Anderson, L.~D.\ 2011, \apj, 738, 27 
\bibitem[Battersby et al.(2011)]{bat11} Battersby, C., Bally, J., Ginsburg, A., et al.\ 2011, \aap, 535, A128 
\bibitem[Beaumont \& Williams(2010)]{be10} Beaumont, C.~N., \& Williams, J.~P.\ 2010, \apj, 709, 791 
\bibitem[Benjamin et al.(2003)]{be03} Benjamin, R.~A., Churchwell, E., Babler, B.~L., et al.\ 2003, \pasp, 115, 953 
\bibitem[Bertoldi(1989)]{be89} Bertoldi, F.\ 1989, \apj, 346, 735 
\bibitem[Billot et al.(2010)]{bi10} Billot, N., Noriega-Crespo, A., Carey, S., et al.\ 2010, \apj, 712, 797
\bibitem[Bonatto \& Bica(2009)]{bo09} Bonatto, C., \& Bica, E.\ 2009, \mnras, 392, 483
\bibitem[Cardelli et al.(1989)]{ca89} Cardelli, J.~A., Clayton, G.~C., \& Mathis, J.~S.\ 1989, \apj, 345, 245
\bibitem[Carey et al.(2009)]{ca09} Carey, S.~J., Noriega-Crespo, A., Mizuno, D.~R., et al.\ 2009, \pasp, 121, 76 
\bibitem[Churchwell et al.(2006)]{ch06} Churchwell, E., Povich, M.~S., Allen, D., et al.\ 2006, \apj, 649, 759 
\bibitem[Clemens(1985)]{cl85} Clemens, D.~P.\ 1985, \apj, 295, 422 
\bibitem[Dale et al.(2012)]{da12} Dale, J.~E., Ercolano, B., \& Bonnell, I.~A.\ 2012, \mnras, 427, 2852 
\bibitem[Deharveng et al.(2005)]{de05} Deharveng, L., Zavagno, A., \& Caplan, J.\ 2005, \aap, 433, 565 
\bibitem[Deharveng et al.(2008)]{de08} Deharveng, L., Lefloch, B., Kurtz, S., et al.\ 2008, \aap, 482, 585 
\bibitem[Deharveng et al.(2012)]{de12} Deharveng, L., Zavagno, A., Anderson, L.~D., et al.\ 2012, \aap, 546, A74 
\bibitem[Elmegreen \& Lada(1977)]{el77} Elmegreen, B.~G., \& Lada, C.~J.\ 1977, \apj, 214, 725
\bibitem[Elmegreen(2011)]{el11} Elmegreen, B.~G.\ 2011, EAS Publications Series, 51, 45 
\bibitem[Fukuda \& Hanawa(2000)]{fu00} Fukuda, N., \& Hanawa, T.\ 2000, \apj, 533, 911 
\bibitem[Gritschneder et al.(2010)]{gr10} Gritschneder, M., Burkert, A., Naab, T., \& Walch, S.\ 2010, \apj, 723, 971 
\bibitem[Groenewegen(2006)]{gr06} Groenewegen, M.~A.~T.\ 2006, \aap, 448, 181 
\bibitem[Gutermuth et al.(2008)]{gu08} Gutermuth, R.~A., Myers, P.~C., Megeath, S.~T., et al.\ 2008, \apj, 674, 336 
\bibitem[Gutermuth et al.(2009)]{gu09} Gutermuth, R.~A., Megeath, S.~T., Myers, P.~C., et al.\ 2009, \apjs, 184, 18 
\bibitem[Gutermuth et al.(2011)]{gu11} Gutermuth, R.~A., Pipher, J.~L., Megeath, S.~T., et al.\ 2011, \apj, 739, 84 
\bibitem[Herwig(2005)]{he05} Herwig, F.\ 2005, \araa, 43, 435 
\bibitem[Indebetouw et al.(2005)]{in05} Indebetouw, R., Mathis, J.~S., Babler, B.~L., et al.\ 2005, \apj, 619, 931 
\bibitem[Indebetouw et al.(2007)]{in07} Indebetouw, R., Robitaille, T.~P., Whitney, B.~A., et al.\ 2007, \apj, 666, 321 
\bibitem[Jackson et al.(2006)]{ja06} Jackson, J.~M., et al.\ 2006, \apjs, 163, 145 
\bibitem[Kang et al.(2009)]{ka09} Kang, M., Bieging, J.~H., Kulesa, C.~A., \& Lee, Y.\ 2009, \apj, 701, 454 
\bibitem[Kendrew et al.(2012)]{ke12} Kendrew, S., Simpson, R., Bressert, E., et al.\ 2012, \apj, 755, 71 
\bibitem[Kennicutt(1998)]{ke98} Kennicutt, R.~C., Jr.\ 1998, \apj, 498, 541 
\bibitem[Koenig et al.(2008)]{ko08} Koenig, X.~P., Allen, L.~E., Gutermuth, R.~A., et al.\ 2008, \apj, 688, 1142 
\bibitem[Koenig et al.(2012)]{ko12} Koenig, X.~P., Leisawitz, D.~T., Benford, D.~J., et al.\ 2012, \apj, 744, 130 
\bibitem[Krumholz \& McKee(2008)]{kr08} Krumholz, M.~R., \& McKee, C.~F.\ 2008, \nat, 451, 1082 
\bibitem[Kurucz(1993)]{ku93} Kurucz, R.~L.\ 1993, Kurucz CD-ROM, Cambridge, MA: Smithsonian Astrophysical Observatory, |c1993, December 4, 1993
\bibitem[Lawrence et al.(2007)]{la07} Lawrence, A., Warren, S.~J., Almaini, O., et al.\ 2007, \mnras, 379, 1599 
\bibitem[Lefloch \& Lazareff(1994)]{le94} Lefloch, B., \& Lazareff, B.\ 1994, \aap, 289, 559 
\bibitem[Lucas et al.(2008)]{lu08} Lucas, P.~W., Hoare, M.~G., Longmore, A., et al.\ 2008, \mnras, 391, 136 
\bibitem[Maia et al.(2010)]{ma10} Maia, F.~F.~S., Corradi, W.~J.~B., \& Santos, J.~F.~C., Jr.\ 2010, \mnras, 407, 1875 
\bibitem[Marcolino et al.(2009)]{marco09} Marcolino, W.~L.~F., Bouret, J.-C., Martins, F., et al.\ 2009, \aap, 498, 837 
\bibitem[Marigo et al.(2008)]{ma08} Marigo, P., Girardi, L., Bressan, A., et al.\ 2008, \aap, 482, 883 
\bibitem[Martins et al.(2005)]{ma05} Martins, F., Schaerer, D., \& Hillier, D.~J.\ 2005, \aap, 436, 1049 
\bibitem[Markwardt(2009)]{mar09} Markwardt, C.~B.\ 2009, Astronomical Data Analysis Software and Systems XVIII, 411, 251 
\bibitem[Matsakis et al.(1976)]{ma76} Matsakis, D.~N., Evans, N.~J., II, Sato, T., \& Zuckerman, B.\ 1976, \aj, 81, 172 
\bibitem[Mercer et al.(2005)]{me05} Mercer, E.~P., Clemens, D.~P., Meade, M.~R., et al.\ 2005, \apj, 635, 560 
\bibitem[Molinari et al.(2010)]{mo10a} Molinari, S., Swinyard, B., Bally, J., et al.\ 2010a, \pasp, 122, 314 
\bibitem[Molinari et al.(2010)]{mo10b} Molinari, S., Swinyard, B., Bally, J., et al.\ 2010b, \aap, 518, L100 
\bibitem[Ossenkopf \& Henning(1994)]{os94} Ossenkopf, V., \& Henning, T.\ 1994, \aap, 291, 943 
\bibitem[Povich et al.(2009)]{po09} Povich, M.~S., Churchwell, E., Bieging, J.~H., et al.\ 2009, \apj, 696, 1278 
\bibitem[Povich \& Whitney(2010)]{po10} Povich, M.~S., \& Whitney, B.~A.\ 2010, \apjl, 714, L285 
\bibitem[Povich et al.(2011)]{po11} Povich, M.~S., Smith, N., Majewski, S.~R., et al.\ 2011, \apjs, 194, 14 
\bibitem[Rathborne et al.(2006)]{ra06} Rathborne, J.~M., Jackson, J.~M., \& Simon, R.\ 2006, \apj, 641, 389 
\bibitem[Rathborne et al.(2010)]{ra10} Rathborne, J.~M., Jackson, J.~M., Chambers, E.~T., et al.\ 2010, \apj, 715, 310 
\bibitem[Reach et al.(2004)]{re04} Reach, W.~T., Rho, J., Young, E., et al.\ 2004, \apjs, 154, 385 
\bibitem[Rho et al.(2006)]{rh06} Rho, J., Reach, W.~T., 
Lefloch, B., \& Fazio, G.~G.\ 2006, \apj, 643, 965 
\bibitem[Robitaille et al.(2006)]{ro06} Robitaille, T.~P., Whitney, B.~A., Indebetouw, R., Wood, K., \& Denzmore, P.\ 2006, \apjs, 167, 256 
\bibitem[Robitaille et al.(2007)]{ro07} Robitaille, T.~P., Whitney, B.~A., Indebetouw, R., \& Wood, K.\ 2007, \apjs, 169, 328 
\bibitem[Robitaille et al.(2008)]{ro08} Robitaille, T.~P., Meade, M.~R., Babler, B.~L., et al.\ 2008, \aj, 136, 2413
\bibitem[Roman-Duval et al.(2010)]{ro10} Roman-Duval, J., Jackson, J.~M., Heyer, M., Rathborne, J., \& Simon, R.\ 2010, \apj, 723, 492 
\bibitem[Salpeter(1955)]{sa55} Salpeter, E.~E.\ 1955, \apj, 121, 161 
\bibitem[Samal et al.(2012)]{sa12} Samal, M.~R., Pandey, A.~K., Ojha, D.~K., et al.\ 2012, \apj, 755, 20 
\bibitem[Siess et al.(2000)]{si00} Siess, L., Dufour, E., \& Forestini, M.\ 2000, \aap, 358, 593 
\bibitem[Simpson et al.(2012)]{si12} Simpson, R.~J., Povich, M.~S., Kendrew, S., et al.\ 2012, \mnras, 424, 2442 
\bibitem[Simon et al.(2001)]{si01} Simon, R., Jackson, J.~M., Clemens, D.~P., Bania, T.~M., \& Heyer, M.~H.\ 2001, \apj, 551, 747 
\bibitem[Skrutskie et al.(2006)]{sk06} Skrutskie, M.~F., Cutri, R.~M., Stiening, R., et al.\ 2006, \aj, 131, 1163 
\bibitem[Solomon et al.(1987)]{so87} Solomon, P.~M., Rivolo, A.~R., Barrett, J., \& Yahil, A.\ 1987, \apj, 319, 730 
\bibitem[Stil et al.(2006)]{st06} Stil, J.~M., Taylor, A.~R., Dickey, J.~M., et al.\ 2006, \aj, 132, 1158 
\bibitem[Tafalla et al.(2004)]{ta04} Tafalla, M., Myers, P.~C., Caselli, P., \& Walmsley, C.~M.\ 2004, \aap, 416, 191 
\bibitem[Testi et al.(1997)]{te97} Testi, L., Palla, F., Prusti, T., Natta, A., \& Maltagliati, S.\ 1997, \aap, 320, 159 
\bibitem[Testi et al.(1999)]{te99} Testi, L., Palla, F., \& Natta, A.\ 1999, \aap, 342, 515 
\bibitem[Thompson et al.(2012)]{th12} Thompson, M.~A., Urquhart, J.~S., Moore, T.~J.~T., \& Morgan, L.~K.\ 2012, \mnras, 421, 408 
\bibitem[Urquhart et al.(2008)]{ur08} Urquhart, J.~S., Busfield, A.~L., Hoare, M.~G., et al.\ 2008, \aap, 487, 253 
\bibitem[Watson et al.(2008)]{wa08} Watson, C., Povich, M.~S., Churchwell, E.~B., et al.\ 2008, \apj, 681, 1341 
\bibitem[Weaver et al.(1977)]{we77} Weaver, R., McCray, R., Castor, J., Shapiro, P., \& Moore, R.\ 1977, \apj, 218, 377 
\bibitem[Whitney et al.(2003)]{wh03} Whitney, B.~A., Wood,K., Bjorkman, J.~E., \& Cohen, M.\ 2003, \apj, 598, 1079 
\bibitem[Whitney et al.(2008)]{wh08} Whitney, B.~A., Sewilo, M., Indebetouw, R., et al.\ 2008, \aj, 136, 18 
\end{thebibliography}
\end{document}